\documentclass[12pt]{article}
\usepackage{epsfig}
\usepackage{axodraw}

\newcommand{\mrm}[1]{\mathrm{#1}}

\newcommand{\Je}{{\sc{Jetset}}}

\newcommand{\Py}{{\sc{Pythia}}}
 
\newcommand{\cost}{\cos\theta}

\newcommand{\pT}{p_{\perp}}

\newcommand{\sumpT}{\sum{\pT}}
 
\renewcommand{\b}{\mrm{b}}
\renewcommand{\c}{\mrm{c}}
\renewcommand{\d}{\mrm{d}}
\newcommand{\e}{\mrm{e}}

\newcommand{\g}{\mrm{g}}

\newcommand{\nbar}{\overline{\mrm{n}}}

\newcommand{\q}{\mrm{q}}
\newcommand{\s}{\mrm{s}}

\renewcommand{\u}{\mrm{u}}

\newcommand{\BB}{\mrm{B}}

\newcommand{\W}{\mrm{W}}
\newcommand{\Z}{\mrm{Z}}

\newcommand{\cbar}{\overline{\mrm{c}}}
\newcommand{\dbar}{\overline{\mrm{d}}}

\newcommand{\pbar}{\overline{\mrm{p}}}
\newcommand{\qbar}{\overline{\mrm{q}}}

\newcommand{\ubar}{\overline{\mrm{u}}}

\newcommand{\red}{\mrm{r}}
\newcommand{\blue}{\mrm{b}}
\newcommand{\green}{\mrm{g}}
\newcommand{\redbar}{\overline{\mrm{r}}}
\newcommand{\bluebar}{\overline{\mrm{b}}}
\newcommand{\greenbar}{\overline{\mrm{g}}}

\newcommand{\Jpsi}{\mrm{J}/\psi}
\newcommand{\ee}{\e^+\e^-}

\newcommand{\qqbar}{\q\qbar}

\newcommand{\Znoll}{\Z^0}
\newcommand{\gammaZ}{\gamma^* / \Znoll}

\newenvironment{Itemize}{\begin{list}{$\bullet$}%
{\setlength{\topsep}{0.2mm}\setlength{\partopsep}{0.2mm}%
\setlength{\itemsep}{0.2mm}\setlength{\parsep}{0.2mm}}}%
{\end{list}}
\newcounter{enumct}
\newenvironment{Enumerate}{\begin{list}{\arabic{enumct}.}%
{\usecounter{enumct}\setlength{\topsep}{0.2mm}%
\setlength{\partopsep}{0.2mm}\setlength{\itemsep}{0.2mm}%
\setlength{\parsep}{0.2mm}}}{\end{list}}
 
\newlength{\captivewidth}
\newlength{\abstwidth}
\def\pl#1#2#3    {{\em Phys. Lett.} {\bf#1} (#2) #3}
\def\prep#1#2#3  {{\em Phys. Rep.} {\bf#1} (#2)~#3}
\def\prev#1#2#3  {{\em Phys. Rev.} {\bf#1} (#2) #3}
\def\prl#1#2#3   {{\em Phys. Rev. Lett.} {\bf#1} (#2) #3}
\def\zp#1#2#3    {{\em Z.Phys.} {\bf#1} (#2) #3}

\begin{document}

\sloppy

\pagestyle{empty}

\begin{flushright}
LU TP 96--15\\
June 1996\\
\end{flushright}

\vspace{\fill}

\begin{center}
{\LARGE\bf Effects of colour reconnection\\[4mm]
 in W$^{\mathbf{+}}$ W$^{\mathbf{-}}$ events}\\[10mm]

{\Large Emanuel Norrbin\footnote{emanuel@thep.lu.se}\\[6mm]}

Department of Theoretical Physics, University of Lund, \\
S\"olvegatan 14A, S-223 62 Lund, Sweden\\[5mm]

Master Thesis 20p\\
Thesis advisor: Torbj\"orn Sj\"ostrand\footnote{torbjorn@thep.lu.se}\\
\end{center}
 
\vspace{\fill}
 
\begin{center}
{\bf Abstract}\\[2mm]
\noindent
\begin{minipage}{\abstwidth}
We have investigated some observable consequences of colour reconnection in the
reaction
$\ee \rightarrow \W^+\W^- \rightarrow \q_1\qbar_2\q_3\qbar_4$. We use a measure
based on the
momentum structure of the event, comparing reconnected events with `standard'
ones.
Some different recoupling models are described and studied as well as simpler
toy models
to show where the effects are present and what makes most of them disappear in
realistic
models. Some attempts are made to introduce suitable cuts in order to select 
the interesting events.
\end{minipage}
\end{center}
 
\vspace{\fill}

\noindent
\clearpage
\pagestyle{plain}
\setcounter{page}{1} 

\section{Introduction}
\label{sec:introduction}

The reaction $\ee \rightarrow \gammaZ \rightarrow \q\qbar$ is well known and has
been
studied for some time, for example at LEP at CERN, where the $\Znoll$ has been
produced
at roughly 90 GeV energy. This reaction is well described by QCD in the
perturbative (shower) region and the 
Lund string model~\cite{Lundmod} in the non perturbative (fragmentation) one.
In the Lund model strings are drawn between the quarks, possibly via radiated
gluons.
The strings can be seen as colour flux tubes or vortex lines.
The subsequent fragmentation of a string will produce the observed particles.
Events generated
in this manner with the {\Je} and {\Py} programs~\cite{Manual} have successfully
described
many observed phenomena. A summary of the experience from LEP1 is given in
\cite{QCDEventGenerators},
where a discussion of the extrapolation to LEP2 energies is also given.
The reaction of primary interest to us in this work is:
\begin{equation}
\label{eq:etowtoq}
\ee \rightarrow \W^+\W^- \rightarrow \q_1\qbar_2\q_3\qbar_4~.
\end{equation}
One of the goals of LEP2 is to study this reaction experimentally from
the threshold at 160 GeV onwards. This experiment starts this year and
will go on for the rest of the century. Some ten thousand events should be
observed at LEP2, but
not all will be of interest to us, so the low statistics will be a big problem
when we want to apply our methods on future experimental data.
We are for example not interested in events that
produce leptons, and we also need to introduce certain cuts that further limit
the number of surviving events.

A first attempt to describe the reaction~(\ref{eq:etowtoq}) would be to
assume that the $\qqbar$-pair from $\W^+$ forms
one colour singlet and that the $\qqbar$-pair from $\W^-$ forms a second one,
and then the two systems
shower and fragment independently of each other like two overlapping $\Znoll$
decays at LEP.
The $\W$-particles, however, exist only for a very short time, therefore the
space--time
separation between the production points of the two $\q\qbar$-pairs
is very small. In one extreme case you could assume that all four quarks are
produced at the same point.
Thus, in addition to the original colour dipoles $\q_1\qbar_2$ and
$\q_3\qbar_4$,
it would be possible to form another set of
dipoles, namely $\q_1\qbar_4$ and $\q_2\qbar_3$. In the framework of the 
Lund model this would mean that the fragmenting strings should
be drawn differently depending on what quarks form the dipoles. It is not
unreasonable 
to assume that this should
give observable effects in the final state.

The possibility of colour reconnection  has been studied before, for example in
\cite{GPZ} where a simple
`instantaneous' reconnection scenario is discussed. The rearranged dipoles are
here supposed to form immediately
and then shower and fragment independently of each other. Under ideal
geometrical conditions
the effects can be very large, but it has been shown that the distance
between the decay vertices of the W-particles is large
enough for perturbative QCD-radiation in the original dipoles
to occur independently of each other in a first approximation~\cite{SjoValery};
therefore
the instantaneous scenario is not very realistic.
The non-perturbative phase, however, extends much further in space so here the
possibility 
of colour rearrangement is much larger. A summary of the different models is
given in \cite{reconnectionsummary}.

In~\cite{SjoValery} several more or less realistic models for non-perturbative
reconnection have been proposed.
These will be described in more detail in section~\ref{sec:theory} and are those
primarily
studied in this work. Gustafson and H\"akkinen \cite{GJari}
have another approach where they draw the strings in such a way that the 
potential energy of the strings (lambda-measure) is
minimized. In this way differences between reconnected and `normal' events can
be seen, barely, in
the rapidity distribution of charged particles.
Since this signal is indirect, we here propose a method which will allow
a more direct study of the string topology.

In this work we will study an observable constructed from the momentum of the
hadrons in
the final state. Well-behaved four-jet events are studied, the jets
paired two by two in all possible ways and each pair is boosted to the system
where
the jet pairs are back-to-back. In this system the sum of the transverse momenta
of the particles belonging to the pair is calculated. If there is
a string between the quarks supposed to be mothers of these jets the sum should
be
minimized. The method will be discussed further in sections~\ref{sec:theory}
and~\ref{sec:method} where it will be applied to three-jet and four-jet events
in $\Znoll$ decays.

The main reasons to study the effects of colour reconnection
are the possibility to determine the structure of
the QCD vacuum and to find possible systematic errors in the measurement of the
mass
of the $\W$ particle. Models based on superconductivity have been proposed in
e.g. \cite{GPZ}
and \cite{SjoValery}, where the strings are supposed to behave like either Type
I or
Type II superconductors. The goal is to study different distributions for the
reconnection models and find differences between these. In the end the
distributions  could
be compared to experiment in order to either verify or reject the models.
In this work small observable differences have been discovered in
the more realistic models, but we also study some simpler
toy models and note that the effects can be quite significant. We therefore
analyze 
where the effects diminish.

\section{Theoretical overview}
\label{sec:theory}

\begin{figure}
\begin{center}
\mbox{\epsfig{file=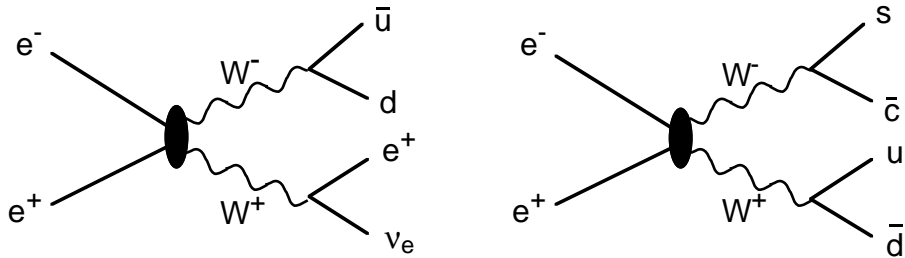}}
\end{center}
\caption[]{Some examples of the reaction $\ee \rightarrow \W^+\W^- \rightarrow$%
leptons and quarks.}\label{fig:reactionexamples}
\end{figure}

Let us study the reaction~(\ref{eq:etowtoq}) in a little more detail to better
understand the different
reconnection models. The process can roughly be divided into the following
stages:
\begin{Itemize}
\item $\ee$ annihilation and $\W^+\W^-$ production;
\item weak $\W^+\W^-$ decay to quarks and leptons; 
\item parton shower;
\item fragmentation (hadronization);
\item unstable particles decay into the observed particles.
\end{Itemize}
The annihilation of electrons and positrons is a very clean process in the sense
that both the initial and final states consist of structureless particles.
The experimental situation is also simplified by the fact that the
reaction takes place in the laboratory rest-frame.
Figure~\ref{fig:reactionexamples} gives some examples of possible reactions. The
black blobs
are there to emphasize that several graphs contribute to the matrix element for
this reaction.

\subsection{No reconnection scenario}

The W-particles live for a very short time and the separation between their
decay points
is less than 0.1~fm ($10^{-16}$ m). The W boson has several decay modes. The
only one that
interests us is $\W\rightarrow\q\qbar$.
Assuming quark--lepton symmetry, no quark mixing, and the fact that the
top-quark is much heavier
than the $\W$-particle, approximately 45\% of the $\W^+\W^-$ events produce two
quark--antiquark pairs. 
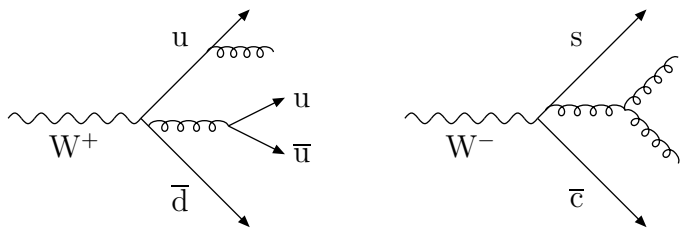
\begin{figure}
\begin{center}
\begin{picture}(300,90)(0,0)
\Photon(0,50)(50,50){2}{5}
\LongArrow(50,50)(90,90)
\LongArrow(50,50)(90,10)
\Text(25,40)[]{$\W^+$}
\Text(65,80)[]{$\u$}
\Text(65,20)[]{$\dbar$}
\Gluon(75,75)(100,75){2}{4}
\SetOffset(3,-3)
\Gluon(50,50)(80,50){2}{4}
\LongArrow(80,50)(100,60)
\LongArrow(80,50)(100,40)
\Text(108,60)[]{$\u$}
\Text(108,40)[]{$\ubar$}
\SetOffset(150,0)
\Photon(0,50)(50,50){2}{5}
\LongArrow(50,50)(90,90)
\LongArrow(50,50)(90,10)
\Text(25,40)[]{$\W^-$}
\Text(65,80)[]{$\s$}
\Text(65,20)[]{$\cbar$}
\SetOffset(153,3)
\Gluon(50,50)(80,50){2}{4}
\Gluon(80,50)(100,70){2}{4}
\Gluon(80,50)(100,30){2}{4}
\end{picture}
\end{center}
\caption[]{Simplified parton showers. Curly lines are gluons.}\label{fig:shower}
\end{figure}

Then each quark-antiquark pair is allowed to radiate other quarks and gluons via
the basic
branchings $\q\rightarrow\q\g$, $\g\rightarrow\g\g$, and $\g\rightarrow\q\qbar$.
In this region the particles are highly virtual and the distances are very small
(less than 1~fm),
so perturbative QCD is applicable. This shower is cut off at some lower 
virtuality scale $Q_0 \approx$ 1~GeV.
A simplified shower is shown in figure~\ref{fig:shower}. It will turn out that
this stage
blurs out most of the effects of colour reconnection.

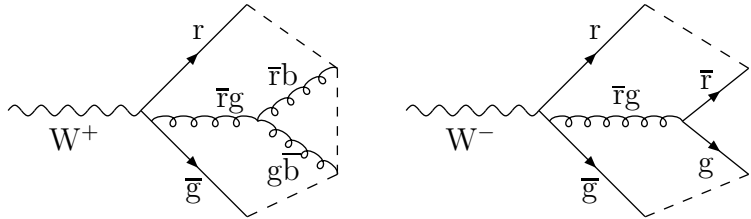
\begin{figure}
\begin{center}
\begin{picture}(300,100)(0,0)
\Photon(0,50)(50,50){2}{5}
\ArrowLine(50,50)(90,90)
\ArrowLine(50,50)(90,10)
\Text(25,40)[]{$\W^+$}
\Text(72,80)[]{$\red$}
\Text(70,20)[]{$\greenbar$}
\SetOffset(4,-4)
\Gluon(50,50)(90,50){2}{4}
\Gluon(90,50)(120,70){2}{4}
\Gluon(90,50)(120,30){2}{4}
\Text(80,58)[]{$\redbar\green$}
\Text(100,68)[]{$\redbar\blue$}
\Text(100,32)[]{$\green\bluebar$}
\DashLine(120,70)(120,30){4}
\DashLine(86,94)(120,70){4}
\DashLine(86,14)(120,30){4}
\SetOffset(0,0)
\Photon(150,50)(200,50){2}{5}
\ArrowLine(200,50)(240,90)
\ArrowLine(200,50)(240,10)
\Text(175,40)[]{$\W^-$}
\Text(222,80)[]{$\red$}
\Text(220,20)[]{$\greenbar$}
\SetOffset(4,-4)
\Gluon(200,50)(250,50){2}{6}
\Text(230,59)[]{$\redbar\green$}
\ArrowLine(250,50)(275,70)
\ArrowLine(250,50)(275,30)
\Text(260,66)[]{$\redbar$}
\Text(260,36)[t]{$\green$}
\DashLine(236,94)(275,70){4}
\DashLine(236,14)(275,30){4}
\end{picture}
\end{center}
\caption[]{Example of string configuration. Only colours are shown in the %
picture, %
r=red, g=green, b=blue. Dashed lines are strings.}\label{fig:strings}
\end{figure}

When the quarks are more than about 1~fm apart perturbation theory breaks down
and
different models are used to describe the fragmentation process. We use the
standard Lund model
where strings are drawn between partons with opposite colour charges.
Figure~\ref{fig:strings}
shows some typical string configurations and the colours of the different
partons. The
term `parton' is a collective term for quarks and gluons. The string could  be
seen
as a force field that is essentially one-dimensional. This arises from the
unique properties of QCD where
the gluon self coupling makes the field lines want to stick together, unlike QED
where
the field lines spread throughout all space. In such a one-dimensional field the
force
would not depend on the separation  between the partons, so the energy in the
colour field
increases approximately linearly with the distance between the quarks. Within
this
colour flux tube new $\q\qbar$-pairs can be created from the increasing field
energy. 
This fragmentation goes on until only ordinary hadrons remain. Some of these
hadrons may be unstable and decay further into the stable hadrons, leptons and
photons that are actually observed. $\pT$, the transverse momentum of the
produced 
hadrons relative to the string direction, fluctuates and has an approximately
Gaussian spectrum.

Most of what has been described so far, $\W^+\W^-$ production, quarks, parton
shower, fragmentation,
and decay are never directly observed. What is seen in the
detector is the stable particles produced in the reaction. The number of
observed particles
varies a lot from event to event but it is in the order of 50. The particles are
not distributed isotropically over all angles, instead they are clearly
collected in two or more
narrow sprays of particles. Therefore the concept of jets is introduced, where
particle momentum
is summed up according to different clustering algorithms so as to better
reflect the 
direction of the parent quarks. In order to be able to compare results
with experiment we mainly use observable quantities in our analysis. Sometimes,
however, we use
the benefits of having a Monte Carlo `behind the scene' view to test our
results.

\begin{figure}
\begin{center}
\mbox{\epsfig{file=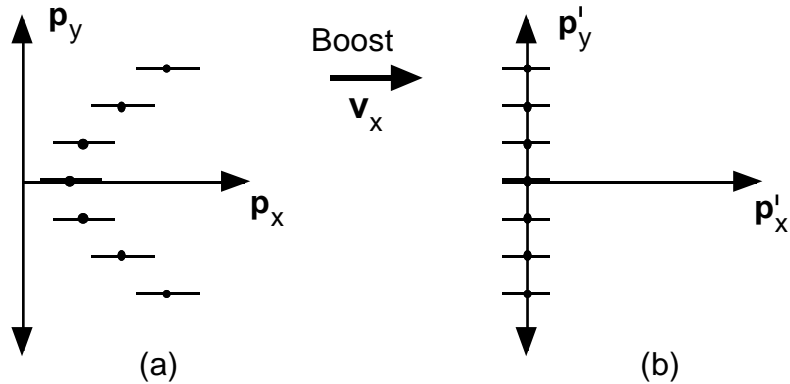}}
\end{center}
\caption[]{Momentum structure of the jets. \textbf{a} Lab system \textbf{b} %
Back-to-back system. %
Fluctuations are represented by the horizontal lines.} \label{fig:jets}
\end{figure}

Consider next the momentum structure of the jets and strings, specifically the
back-to-back system of a jet pair with a string between them.
The momenta
of the individual particles that constitute the jets are not necessarily
parallel to
the direction of the jets but most of them are almost aligned, because the
particles are the remains
of the fragmented string drawn between the $\q\qbar$-pair.
In momentum space the particles in the rest frame are distributed along a
hyperbola as in figure~\ref{fig:jets}(a).
The directions of the quarks are in the asymptotes of the hyperbola.
If the system is boosted to the back-to-back system the momenta of the particles
constituting the jet will be almost aligned
to the jet axis as in (b). In a real
jet the transverse momentum fluctuates slightly and gives a smeared out picture
as is indicated by the lines in the figure. If the quarks belonging
to the back-to-back jets do not have a string drawn between them they will not
follow a
hyperbola and will be more spread out in the back-to-back system. We will
be interested in this broadening in our event analysis.

\subsection{Colour reconnection}
\label{sec:theory:reconnection}

If we imagine that the $\W^+$ and $\W^-$ systems  shower and fragment
independently
of each other we would have no problem. Then the produced hadrons are, in
theory, uniquely assigned
either to the $\W^+$-system or to the $\W^-$-system. The $\W$-particles decay
very
close to each other both in space and time, however, so this kind of separation
might not be possible.
It is therefore useful to discuss the wavelength of gluons in different stages
of the process.
A hard gluon in the perturbative region has a wavelength much smaller than the
decay vertex separation
and can therefore resolve the two decay vertices.
Indeed, it has been shown \cite{SjoValery} that QCD interference effects are
negligible for energetic
gluon emission involving energies significantly larger than
$\Gamma_{\W}$\footnote{The invariant mass of 
a $\W$-particle produced in the reaction~(\ref{eq:etowtoq}) is distributed
according to
a Breit-Wigner distribution with a width, $\Gamma_{\W}$, in the order of 2
GeV.}.
In the fragmentation region, on the other hand, gluon wavelengths are much
larger and
interference effects should be more important.
Having said this we concentrate on the possibility of reconnection in
the non-perturbative fragmentation region where the typical distance a parton
travels before
branching, in the parton shower picture, is larger than the $\W^+\W^-$
separation.

As has been mentioned before, the complexity of QCD prohibits a description of
the fragmentation process from first
principles so that reconnection in this phase has to
be modelled too. Several different models has been proposed and they all build
on different assumptions
about the QCD vacuum. The ones studied in this work are:
\begin{enumerate}
\item Reconnection at origin of event. A simple `instantaneous' model
where strings are always reconnected. Reconnection before shower and
fragmentation. Not considered to be realistic.
\item Same as above but with reconnection after shower but before fragmentation.
The `intermediate' model. The simplest of the more realistic models.
\item Reconnection when strings overlap based on cylindrical geometry. A 
`bag model' based on a type I superconductor. The reconnection probability
is proportional to the overlap integral between the field strengths. The field
strength has a
Gaussian fall-off in the transverse direction, and a radius of about 
0.5 fm. The model contains a free strength parameter that
can be modified to give any reconnection probability.
\item Reconnection when strings cross. In this model the strings mimic the
behaviour of the vortex
lines in a type II superconductor where all topological information is given by
a one-dimensional
region in the core of the string.
\item In this model strings are drawn between partons in such a way that the
`length' is 
minimized. As a measure of this length a so called $\lambda$-measure is used.
This
can also be seen as a measure of the potential energy of the string.
\end{enumerate}
\vspace{2mm}
These models will be referred to by their number or by their name. Models 1
through 4 is described
in great detail in \cite{SjoValery} and the last in \cite{GJari}. Models 1 and 2
will also
be studied without parton showers as toy-models.

Colour reconnection is not limited to $\W^+\W^-$ decays. It has been studied in
other reactions
as well, for example in the decay $\BB\rightarrow\Jpsi+X$ where colour
reconnection is needed to
create the observed $\Jpsi$-particle. The $\BB$-meson contains a $\b$-quark and
the $\Jpsi$ is a
$\c\cbar$ state. In the underlying decay
$\b\rightarrow\c\W^-\rightarrow\c\cbar\s$
the natural colour singlet would be $\s\cbar$ and only in 1/9 of the cases would
$\c\cbar$
be a possible singlet system. Without the possibility of colour reconnection the
production
of $\Jpsi$ would be more suppressed than it has been observed to be.
Other examples are mentioned in~\cite{SjoValery}.

\section{Method of analysis}
\label{sec:method}

In this section we present the method used to study colour reconnection.
We first describe the principle behind the method and then apply it to the
reaction:
\begin{equation}
\label{eq:etoztoq}
\ee\rightarrow\Znoll\rightarrow\q\qbar\rightarrow\q\qbar\g
\end{equation}
where a hard gluon is emitted by the initial $\q\qbar$-pair. This reaction
typically
produces a three-jet event where one jet is associated with the gluon and the
other two
with the quarks. We use this, as a test of our method, to identify the gluon
jet.
A second example involving four-jets is also studied in a simple case.
In section~\ref{sec:main} we proceed with
the study of reaction~(\ref{eq:etowtoq}), where we want to find out how the
strings are drawn using
different methods and different recoupling models.

\subsection{The $\pT$ minimization method}

We consider an event producing a few well separated jets that can be paired
one-to-one with
the same number of initial quarks or hard gluons. Then we consider the possible
string configurations and
try to find out what configuration has been realized in a specific event. In the 
first example that we study, reaction~\ref{eq:etoztoq}, 
there is only three possible string configurations with two strings in each.
Therefore we set up three hypotheses and proceed to find the right one by
looking
at the back-to-back system of every pair of jets that are assumed to have
strings between them.
In this system $\sumpT$---
the sum of the transverse momentum of the particles that `belong' to this jet
pair
--- is calculated. How a particle is assigned to a jet pair depends on the
situation, but in
principle a particle is assigned to the jet pair in whose back-to-back system
its $\pT$
is minimized. This will be made clear in the specific examples. The `right'
hypothesis
is then identified as the one that minimizes $\sumpT$. Why this is so will be
explained
and shown in the following sections. 

\subsection{Test of the method in three-jet events}

\begin{figure}
\begin{center}
\mbox{\epsfig{file=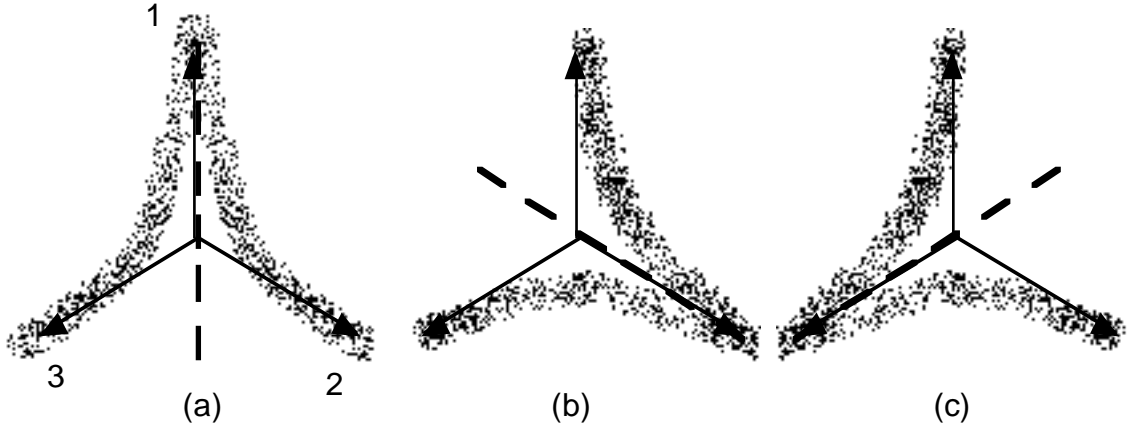}}
\end{center}
\caption[]{The three different hypotheses. The arrows show the direction of %
the jets. %
The gluon is giving rise to: \textbf{a} jet1 %
\textbf{b} jet2 \textbf{c} jet3. The smearing shows the expected momentum %
distribution in %
the respective hypothesis.}
\label{fig:3jet}
\end{figure}

We demonstrate this method by studying the reaction (\ref{eq:etoztoq}) where
the three hypotheses are depicted in
figure~\ref{fig:3jet}. In this case, because of conservation of momentum, the
reaction takes place
in a single plane. This simplifies the analysis somewhat.
In (a) jet~1 is assumed to be the gluon; therefore strings are drawn between
the quarks via jet~1. Alternatively the gluon can be seen as a kink on the
string between the quarks.

All particles to the left of the dashed line are assigned
to the left string and all particles right of the line to the right one. Then
two boosts are performed.
One to the back-to-back system of jet 1 and 2 and one to that of 1 and 3. In
each of these systems
$\sumpT$, for the `right' and `left' particles respectively,
is calculated and the sums are added together. The number obtained in this way
is a measure of the
`goodness' of the hypothesis: 
the smaller the $\sumpT$ is the better particles are lined up along the expected
hyperbola, compare figure~\ref{fig:jets}. The same is done in (b) and (c) but in
these cases the gluon is
assumed to be in the direction of jet 2 and 3 respectively.

Because of the way the momentum of the particles is distributed when a string is
drawn between
jets -- in the ideal case along
a hyperbola -- the right hypothesis should be the one with the smallest
$\sumpT$.
Imagine for example that (a) is the right hypothesis but we make the cut and
boosts according
to (b). When we boost to the 2-3 system, the transverse momenta of particles not
aligned to a 
hyperbola in the rest-frame will be exaggerated and $\pT$ will increase.

We test this method by generating $\Znoll\rightarrow\q\qbar$ events at 91.2 GeV
using the \Je{ }7.4
and \Py{ }5.7 event generators \cite{Manual}. We generate standard events but
switch off initial
state photons and leptonic $\Znoll$ decays.
Using the clustering algorithm in \Je{ }we search for three-jet events and
perform
the analysis described above to identify the gluon jet with our method. In a
Monte
Carlo simulation it is possible to
trace the showering of the primary $\q\qbar$-pair and find the momenta of the
original quarks after the parton shower.
Then we check the direction of the jets and match two of them with the quarks.
The last jet
is assumed to come from a hard gluon. In this way we can see how often we get
the
`right' answer when using this method. 
In a real experimental event this would of course not be possible, other
techniques have
to be used to tag the gluon-jet.
Had the methods been completely uncorrelated we
would be right in 33 \% of the cases. As it turns out we are right in more than
50 \% and
under ideal conditions we get over 80 \%, see table.

For better results we have some parameters that we can change. These are:
\begin{Itemize}
\item $d_{\mrm{join}}$ in the clustering algorithm. This is a distance scale in
GeV above which two
clusters may not be joined. It can be said to control the jet-resolution power.
A higher
number makes the algorithm more prone to join two jets and hence a three-jet
could be
classified as a two-jet if $d_{\mrm{join}}$ is too large. For small numbers it
is the other way around.
\item Particle energy cut. Particles with high energy and hence large momentum
should
not contribute as much to the difference between jet-pairs with and without
strings, since
the string hyperbolae run almost parallel to the jet directions at large
momenta, compare
figure~\ref{fig:3jet}.
\item Jet energy cut. If one of the reconstructed jets has a small energy it
could be
that no hard gluon is associated with this jet and the event should therefore
not be part
of the analysis.
\item The angle between any two jets should be above some minimum value.
This one is not varied but held constant at $60^{\circ}$.
\end{Itemize}
\vspace{2mm}
These parameters are arguably not all independent and it is not our intention to
optimize the cuts,
nevertheless in table~\ref{table:3jet} some results are shown for different
values of
the parameters.
\begin{table}
\begin{center}
\begin{tabular}{|c|c|c|c|c|c|}
\hline
run	& $d_{\mrm{join}}$	& jet-energy	& particle-energy	& \% `right'	& \%
considered	\\
\hline
1	& 3.5		& 8.0		& 2.0			& 58		& 17 		\\
2	& 3.5		& 8.0		& 4.0			& 64		& 17		\\
3	& 3.5		& 8.0		& 10.0			& 63		& 18		\\
\hline
4	& 5.5		& 8.0		& 4.0			& 61		& 23		\\
5	& 5.5		& 12.0		& 4.0			& 58		& 18		\\
6	& 5.5		& 16.0		& 4.0			& 54		& 13		\\
\hline
7	& 7.5		& 8.0		& 4.0			& 57		& 22		\\
8	& 7.5		& 16.0		& 4.0			& 52		& 13		\\
\hline
9	& 2.0		& 12.0		& 4.0			& 57		& 2.0		\\
\hline
10	& 7.5		& 4.0		& 15.0			& 87		& 100		\\
\hline
\end{tabular}
\caption{Run 1--8 and 10: 20,000 generated events. Run 9: 100,000 events and %
optimal cuts, %
a small percent of considered events is not really a problem because of %
the good availability of data. All units are in GeV, jet-energy is lower limit %
while particle-energy is %
upper limit. The last item is results from ideal conditions %
as described in the text.}\label{table:3jet}
\end{center}
\end{table}

Statistics is not really a problem in the case of reaction~(\ref{eq:etoztoq}).
There exists data from 
millions of experimental events at LEP which enables us to study as many events
as we like
in order to get only a small statistical error. In table~\ref{table:3jet}, run
1--6, events
with different cuts are divided into groups of three where only one parameter is
varied. The predictive
power of the method is given by the percentage in column 5. The `\% considered'
should be somewhere around 10\% from the known amount of three-jets in this
reaction. Some conclusions could be drawn from this:
The `\% right' is only moderately sensitive to the cuts in
$d_{\mrm{join}}$, jet-energy and particle-energy.
The cut in jet-energy should be in the larger region to get fewer considered
events.
Row~9 shows `optimized' cuts and has a very low $d_{\mrm{join}}$ (default is 2.5
GeV) and high jet-energy cut.
A low value of $d_{\mrm{join}}$ will force the clustering algorithm to pick out
clear three-jets only and hence
enhance the results.

In a `real life' event generation like in 1--9 of table~\ref{table:3jet}
fluctuations play a
significant part in the results.The gluon-jet, for example, is almost never as
pronounced
as in figure~\ref{fig:3jet} and the jet-axis is not exactly aligned to the
direction of the
quarks and gluons. This will make the alignment of momentum to the hyperbola
uncertain and $\sumpT$ might become larger. In order to see clearly that the
method is correct
we generate events with a $\u\ubar$-pair
and a gluon, each with one third of the total center of mass energy, aligned in
a plane with
equal angles. The result from this is shown in run~10 of table~\ref{table:3jet}.
Here the particles
are collected to begin with so it is better to increase $d_{\mrm{join}}$ to get
as many three-jet events
as possible (in this case 100\%, see table).

\subsection{Four-jet events}

In a three-jet event with planar structure like the one in figure~\ref{fig:3jet}
it has been shown
that the region between the quark-jets contains less particles than the other
two does.
This result is in good agreement with the Lund string fragmentation model but
not
with independent fragmentation. In the independent fragmentation model the
partons
fragment independently of each other and no string effects should be noticed.
This is discussed in~\cite{Lundmod} where experimental as well as theoretical
results
are cited.

\begin{figure}
\begin{center}
\begin{picture}(430,100)(0,0)
\LongArrow(60,50)(100,50)
\Text(110,50)[]{$\u$}
\LongArrow(60,50)(20,50)
\Text(10,50)[]{$\ubar$}
\Gluon(60,50)(60,90){3}{4}
\Text(70,90)[]{$\g$}
\Gluon(60,50)(60,10){3}{4}
\Text(50,10)[]{$\g$}
\Text(60,-10)[]{(a)}
\SetOffset(110,0)
\LongArrow(60,50)(100,50)
\LongArrow(60,50)(20,50)
\LongArrow(60,50)(60,90)
\LongArrow(60,50)(60,10)
\DashLine(100,50)(60,90){4}
\DashLine(20,50)(60,10){4}
\DashLine(60,90)(65,50){4}
\DashLine(65,50)(60,10){4}
\Text(60,-10)[]{(b)}
\SetOffset(220,0)
\LongArrow(60,50)(100,50)
\LongArrow(60,50)(20,50)
\LongArrow(60,50)(60,90)
\LongArrow(60,50)(60,10)
\DashLine(100,50)(60,10){4}
\DashLine(20,50)(60,90){4}
\DashLine(60,90)(65,50){4}
\DashLine(65,50)(60,10){4}
\Text(60,-10)[]{(c)}
\SetOffset(330,0)
\LongArrow(60,50)(100,50)
\LongArrow(60,50)(20,50)
\LongArrow(60,50)(60,90)
\LongArrow(60,50)(60,10)
\DashLine(100,50)(60,10){4}
\DashLine(20,50)(60,90){4}
\Text(60,-10)[]{(d)}
\DashLine(20,50)(60,45){4}
\DashLine(60,45)(100,50){4}
\end{picture}\\
\end{center}
\caption[]{Original parton distribution and possible string configurations %
in a four-jet event. Dashed lines are strings. Strings between the gluons %
have been slightly bent for visibility.}\label{fig:qqgg}
\end{figure}
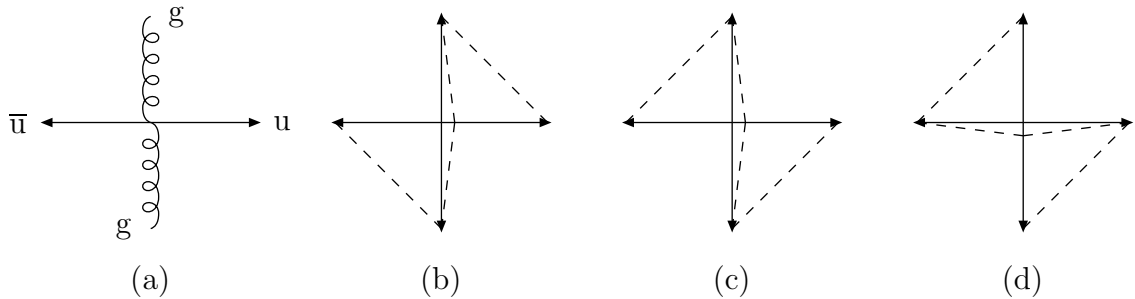

No similar study has been made in four-jet events, no doubt due to the 
difficult geometry. It could be possible to do this by using the
described $\pT$-method. Here we study only a very simple example where
the initial quark-gluon configuration is depicted in figure~\ref{fig:qqgg}a.
This configuration is then allowed to fragment according to the two different
fragmentation models (Lund and independent). Four jets are constructed and in
this case there are twelve
different ways to draw the strings between the four jets. First there are six
ways
to choose which two jets belong to the quark and anti-quark and for each of
these there are two possible
ways to draw the strings via the gluons. Three of these
twelve string configurations are
shown in figure~\ref{fig:qqgg}(b, c and d), where (b) and (c) corresponds to the
parton configuration
in (a) while (d) corresponds to a configuration where the quarks and gluons
change places.
The rule is that each quark must have only one
string attached to it while the gluons must have two. This rule could
lead to other string configurations than those studied here, but these further
ones are colour
suppressed in QCD perturbation theory and are looked upon as second order
corrections. This
kind of colour reconnection, with closed gluon loops, is studied
in~\cite{crister}.

In this case every configuration contains three string pieces, therefore
three boosts to the back-to-back systems of jets with strings
between them are performed and $\sumpT$ is calculated for the twelve hypotheses.
A simple geometrical cut is no longer suitable to
assign a particle to a string, as
was the case for three-jets. Instead $\pT$ is calculated for every particle and
every string system and the particle is assigned to the system where its
$\pT$ is minimized. This is actually equivalent to what was done for three-jets
but easier to generalize to three dimensions.

\begin{figure}
\begin{center}
\mbox{\epsfig{file=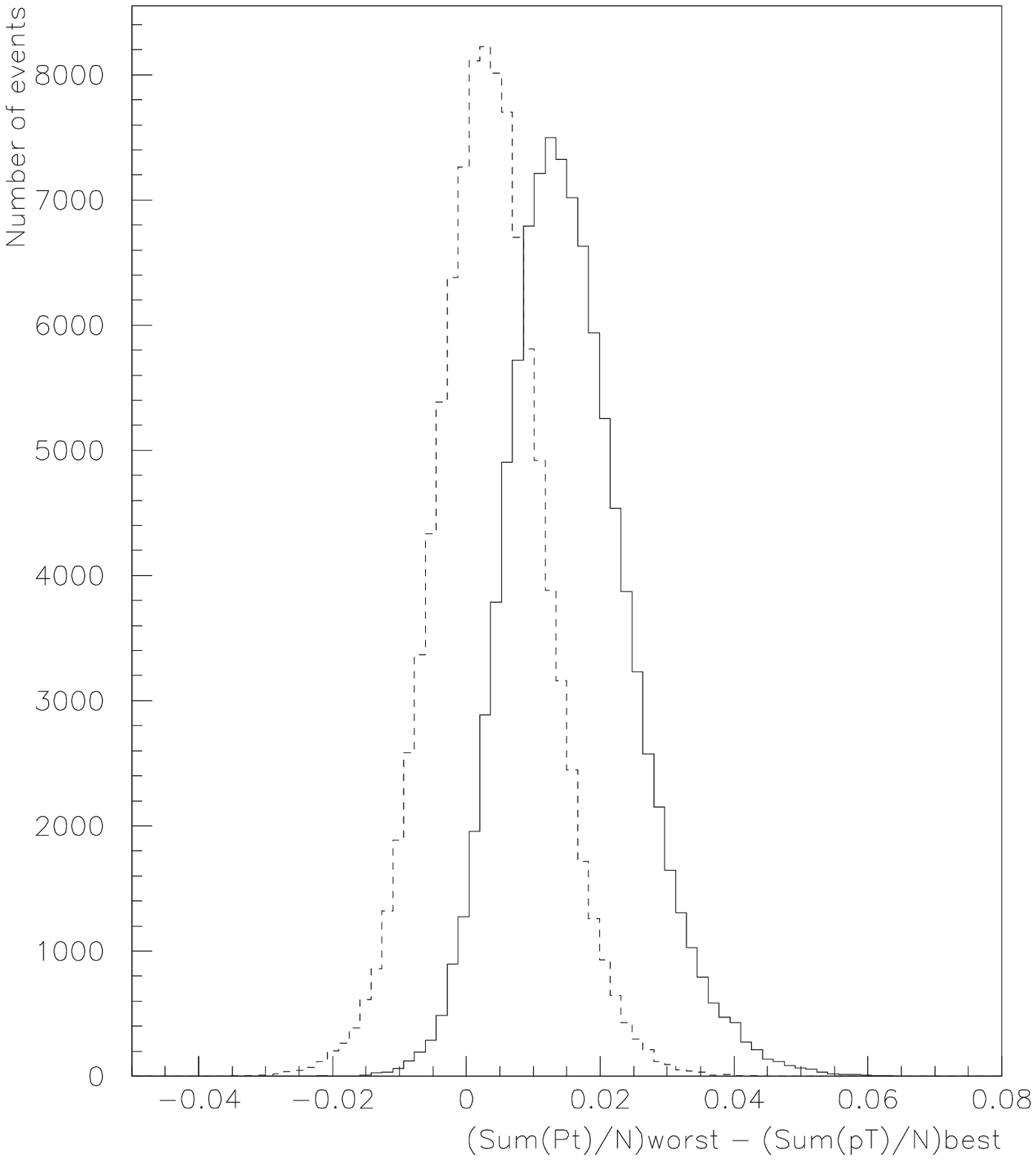, width=70mm, height=70mm}\hspace{10mm}%
\epsfig{file=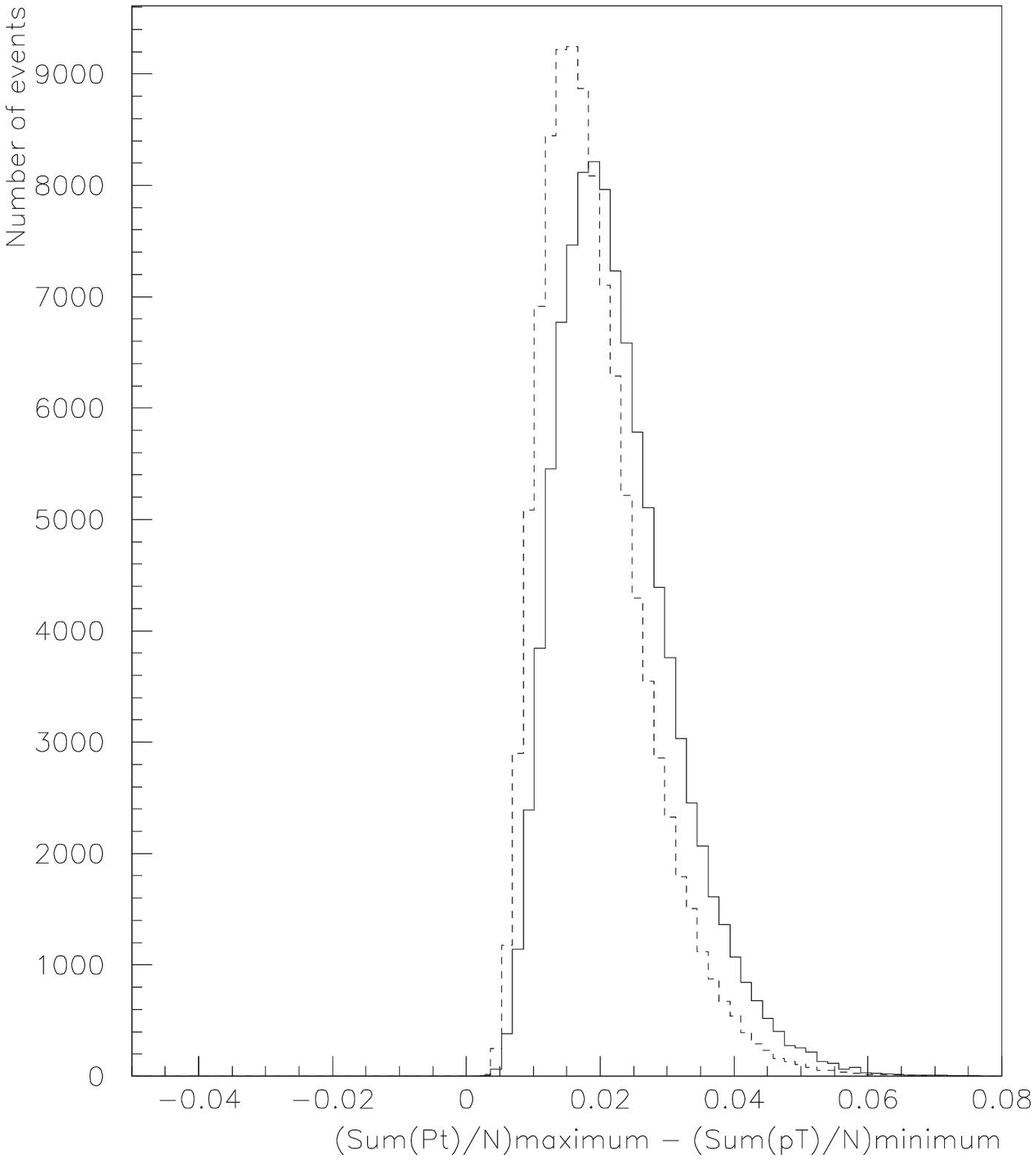, width=70mm, height=70mm}}
\caption[]{The distribution of \textbf{a}: %
$(\sumpT/N)_{\mrm{worst}}-(\sumpT/N)_{\mrm{right}};$ %
and \textbf{b}: $(\sumpT/N)_{\mrm{maximum}}-(\sumpT/N)_{\mrm{minimum}}$ %
for Lund string (full) and %
independent (dashed) fragmentation respectively.}\label{fig:special1}
\end{center}
\end{figure}

Since we have chosen the simple configuration in figure~\ref{fig:qqgg}a from the
beginning we know which of the twelve configurations is the right one
(fig~\ref{fig:qqgg}b)
and which is the worst, where none of the string regions are identified
correctly
(fig~\ref{fig:qqgg}d). Fig~\ref{fig:qqgg}c could also be the correct one but we
choose
(b). We then plot the distribution of the difference,
$(\sumpT/N)_{\mrm{worst}}-(\sumpT/N)_{\mrm{right}}$, see
figure~\ref{fig:special1}a,
for Lund string fragmentation and independent fragmentation respectively. 
Here N is the number of particles in the sum.

For independent fragmentation the distribution is centered at zero
while it is displaced to the positive for Lund string fragmentation.
This shows that the $\pT$-measure is sensitive to the existence of strings.
It could also be used to compare with experiment, in order to either confirm or
reject the string model, but the method would have to
be slightly altered because it is not possible to know which
partons are which in an experimental situation. It is possible, however,
to tag the quarks and be left with two possible string configurations, for
example those in figure~\ref{fig:qqgg}b and c. If we assume that the
$\pT$-method can distinguish these two situation the configuration with
the smallest $\pT$ would be the `right' one. Let us say that this method
chooses b as the correct one, then d would be the worst, where no string
pieces are drawn correctly. We would then, for every event, be left with
a difference, $(\sumpT/N)_{\mrm{worst}}-(\sumpT/N)_{\mrm{right}}$, that
could be plotted in the same way as in figure~\ref{fig:special1}a.

Another possible distribution to study is that of the difference
$(\sumpT/N)_{\mrm{maximum}}-(\sumpT/N)_{\mrm{minimum}}$, where all twelve
hypotheses
are compared without any consideration of which is the best or worst, just
which gives the largest and smallest value of $\sumpT$.
This has been done for the simple topology in figure~\ref{fig:qqgg}a and the
result is shown in figure~\ref{fig:special1}b.
Unfortunately this signal is not large. We would have wanted this figure to show
a larger difference
between the `best' and `worst' scenario in the Lund model as opposed to
independent fragmentation, reflecting
the string effect. Had this been the case the string model could have been
tested in four-jet events
without the need for the quark tagging process. This shows that
fluctuations play a large role in the calculated sums, distorting the simple
expectations.

\section{Colour reconnection in $\W^+\W^-$ events}
\label{sec:main}

We will now proceed with the study of reaction~(\ref{eq:etowtoq}) and colour
reconnection.
Again full events are generated using \Je{} and \Py{}, but now the strings
are allowed to reconnect according to the different models  described in
section~\ref{sec:theory:reconnection}.
The basic idea is to study $\sumpT$ for different string configurations and
compare the
prediction with other methods and with the `real' configuration as generated by
the program.
The `\% right' that we have studied in the three-jet case is just a single
number, but we would like
to have a continuous variable and study the shape of its distribution. A
variable similar to 
the one in the four-jet case will be constructed from $\sumpT$ and its
distribution
will be compared for events with and without reconnection.

The kinds of events that will interest us in this analysis are those where we
have
four distinguishable clusters of particles which are well separated in space.
The
reason for this is that we want to ascribe one jet to each primary quark. To get
the
best results with our method we have to take certain measures:
\begin{Itemize}
\item optimize the cluster finding routine for four-jets;
\item jets must have some minimum energy;
\item the angle between any jet pair should not be too small.
\end{Itemize}
\vspace{2mm}
The angle is fixed at 30$^{\circ}$ to agree with the cuts used
in~\cite{SjoValery}.
This will reduce the number of considered events to about 60 \%. This time
statistics will be a bigger problem because of the small amount of data
that will be available.

\subsection{Some preliminary results}

We will first identify which jet originates from which quark, without 
reconnection, using different methods. These are:
\begin{Enumerate}
\item Looking at the $\q_1\qbar_2\q_3\qbar_4$ configuration before
parton shower and matching these one to one with the reconstructed jets. This is
done by minimizing the products of the four (jet+$\q$) invariant masses. Here we
use
information not available in an experimental situation and it is
used only as a reference.
\item setting $\overline{m}_{\W}^{ }=(m^+_{\W}+m^-_{\W})/2$, where $m_{\W}^{ }$
is the reconstructed
W-masses, minimize $|\overline{m}_{\W}^{ }-80|$.
We already know that the W-mass is about 80 GeV.
\item A variant of the above is to minimize $|m^+_{\W}-80|+|m^-_{\W}-80|$
instead.
\item Using only the spatial direction of the four jets
and maximizing the sum of opening angles between jets from the same pair.
\item In the case of independent shower and fragmentation (no reconnection)
picking the combination that minimizes $\sumpT$, see text.
\end{Enumerate}
\vspace{2mm}

\begin{figure}
\begin{center}
\begin{picture}(300,80)(0,0)
\LongArrow(50,50)(30,20)
\LongArrow(50,50)(30,80)
\LongArrow(50,50)(80,80)
\LongArrow(50,50)(70,30)
\Text(22,20)[]{1}
\Text(22,80)[]{2}
\Text(87,80)[]{3}
\Text(77,30)[]{4}
\Text(50,0)[]{(a)}
\DashLine(30,20)(80,80){4}
\DashLine(30,80)(70,30){4}
\SetOffset(100,0)
\LongArrow(50,50)(30,20)
\LongArrow(50,50)(30,80)
\LongArrow(50,50)(80,80)
\LongArrow(50,50)(70,30)
\Text(50,0)[]{(b)}
\DashLine(30,20)(70,30){4}
\DashLine(30,80)(80,80){4}
\SetOffset(200,0)
\LongArrow(50,50)(30,20)
\LongArrow(50,50)(30,80)
\LongArrow(50,50)(80,80)
\LongArrow(50,50)(70,30)
\Text(50,0)[]{(c)}
\DashLine(30,20)(30,80){4}
\DashLine(80,80)(70,30){4}
\end{picture}
\end{center}
\caption[]{The possible string configurations in a four-jet event. %
Dashed lines are strings. }\label{fig:4jet}
\end{figure}
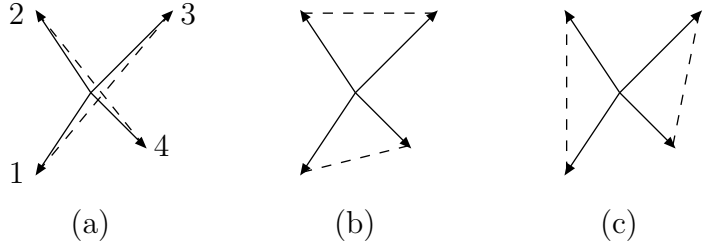

If the $\W$ decays are $\W^+\rightarrow\q_1\qbar_2$ and
$\W^-\rightarrow\q_3\qbar_4$, then in
the case of no reconnection the string configuration is: $\q_1-\qbar_2$ and
$\q_3-\qbar_4$,
possibly with some gluons in between the quarks acting as kinks on the string.
As was the case with three-jets in reaction~(\ref{eq:etoztoq}) 
we have three hypotheses which are depicted in figure~\ref{fig:4jet}.
For each of the three hypotheses 
every particle is assigned to the one of the two
back-to-back systems where its $\pT$ is minimal.
This assignment was also made in the four-jet case. In figure~\ref{fig:4jet} the
respective back-to-back systems are: (a) 1--3, 2--4 (b) 2--3, 1--4 and (c) 1--2,
3--4.
In each back-to-back system $\sumpT$ is calculated and then added to the
$\sumpT$ of
the other system and the configuration with the smallest total sum will then be
identified as the
correct one in method~5.

Ordinary events are generated at 170 GeV and
first we compare method 2--5 with method 1 to see how accurate they are. Then
method 5 is
compared one-to-one with methods 2--4 (see table~\ref{table2}). Of method 2--4
number 3 turned
out to be the most accurate one, so from now on we will only use this one.
\begin{table}
\begin{tabular}{llll}
\hline
No reconnection: \\
Method 1 and method 2	& gives the same result in	& 83	& \% of the cases\\
Method 1 and method 3	& gives the same result in	& 85	& \% of the cases\\
Method 1 and method 4	& gives the same result in	& 83	& \% of the cases\\
\vspace{1mm}
Method 1 and method 5	& gives the same result in	& 68	& \% of the cases\\
Method 5 and method 2	& gives the same result in	& 68	& \% of the cases\\
Method 5 and method 3	& gives the same result in	& 65	& \% of the cases\\
Method 5 and method 4	& gives the same result in	& 64	& \% of the cases\\
\hline
\vspace{1mm}
Reconnected events: \\
Method 1 and method 3	& gives the same result in	& 85	& \% of the cases\\
\vspace{1mm}
Method 1 and method 5	& gives the same result in	& 46	& \% of the cases\\
Method 5 and method 3	& gives the same result in	& 46	& \% of the cases\\
\hline
\end{tabular}
\caption[]{Test of the different methods. 20,000 generated events, with about %
12,000 surviving %
all the cuts. Reconnection according to model~2.}\label{table2}
\end{table}

Next we do the exact same analysis but this time we use reconnected events
according to model
number~2 chapter~\ref{sec:theory:reconnection}, which is simple but still
somewhat realistic. The first three percentages
in table~\ref{table2} are practically unchanged whereas all the percentages for
method~5 ---
the $\sumpT$ method --- are lowered by about 20 percent units. This means that
the first
four methods are not sensitive to how the strings are drawn but method~5 is.
This lowering of the percentage could be compared to experiment in order to
determine the fraction
of reconnected events in a data sample. A simple number, however, will not be
convincing when comparing simulated events with experimental ones because
we don't know if the difference comes from colour reconnection or from some
fault in the model.

\begin{figure}
\begin{center}
\mbox{\epsfig{file=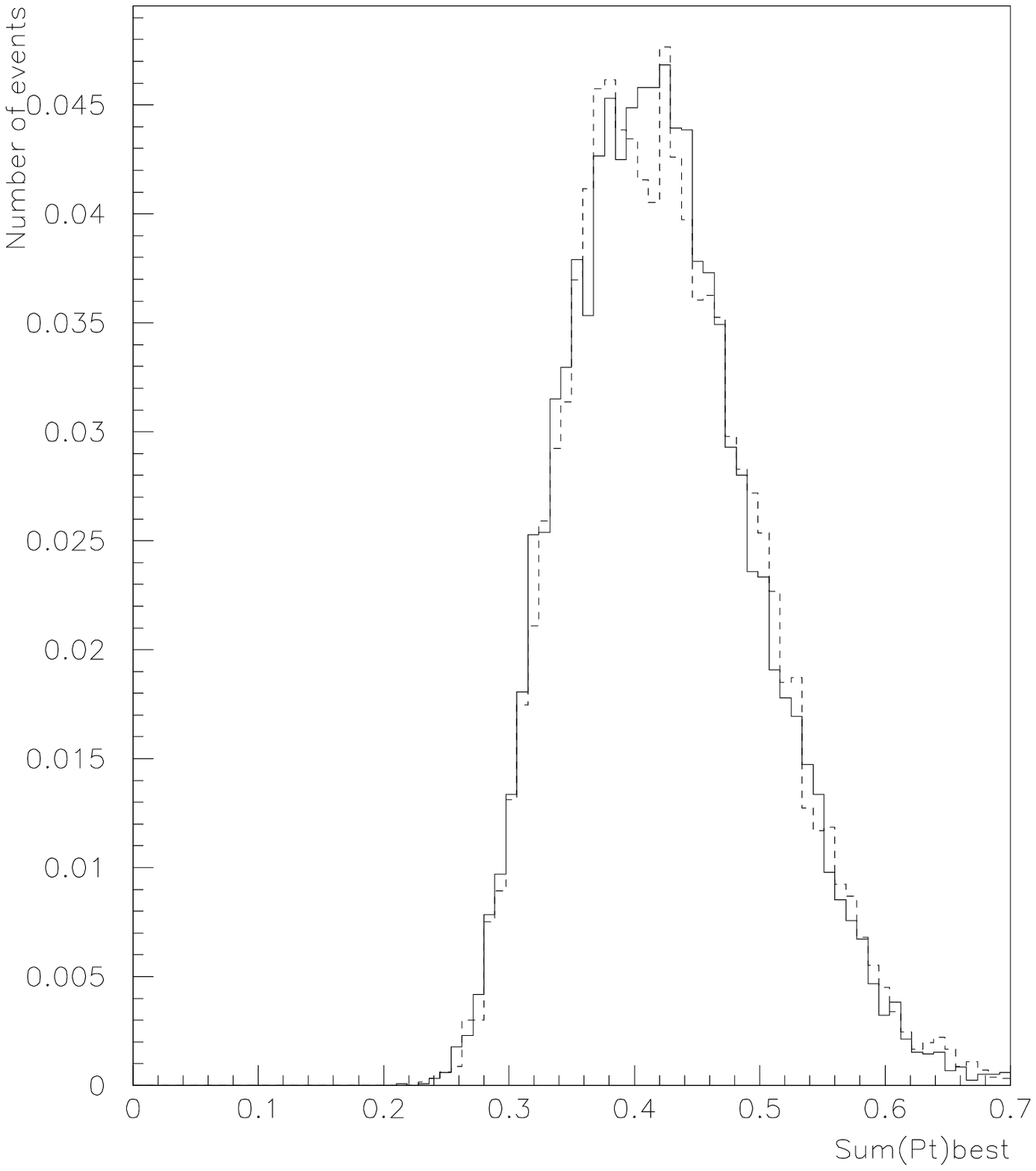, width=60mm, height=60mm}
\hspace{10mm}
\epsfig{file=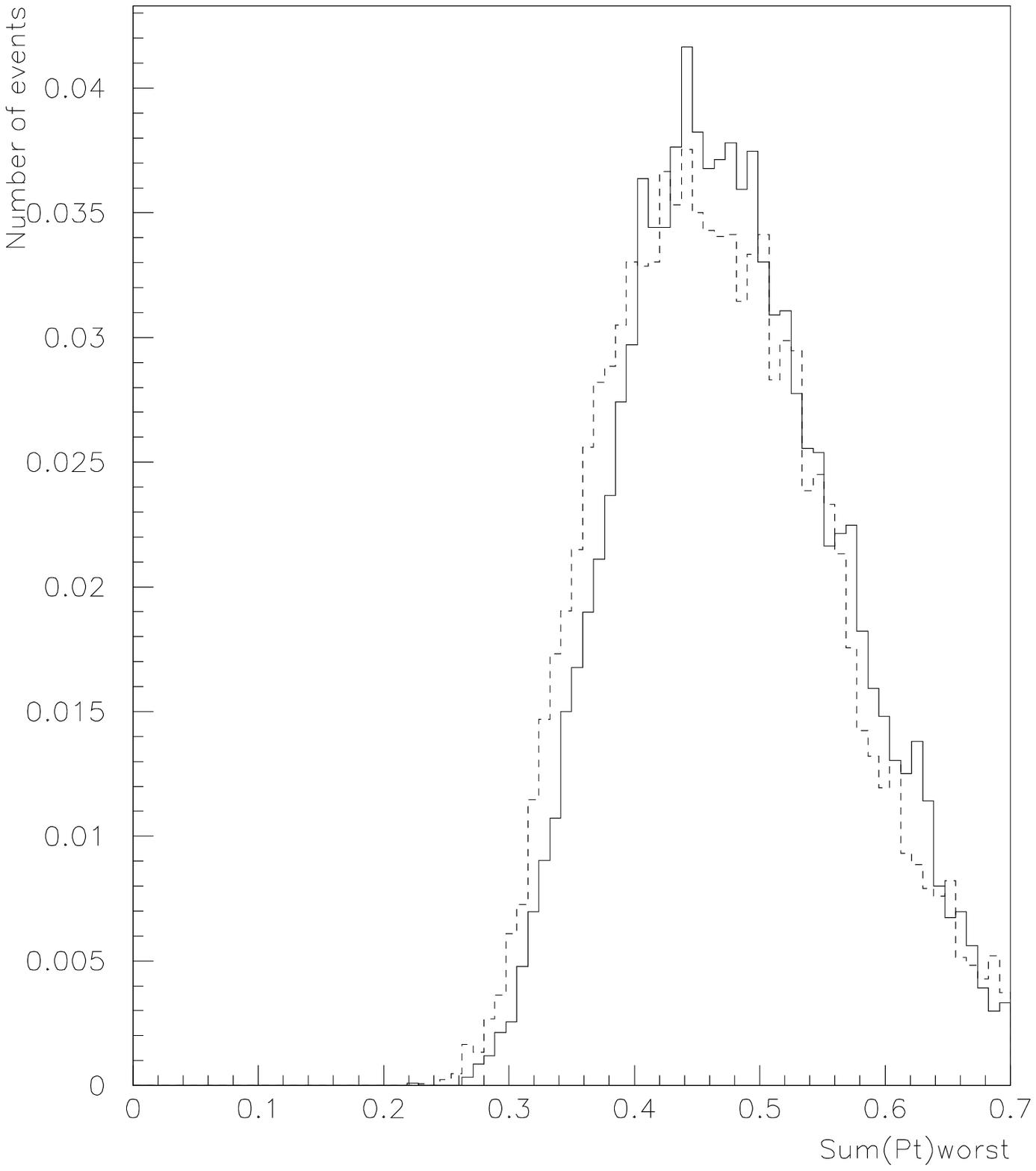, width=60mm, height=60mm}}\\
\vspace{-3mm}\mbox{(a)\hspace{60mm}(b)}\\
\vspace{5mm}
\epsfig{file=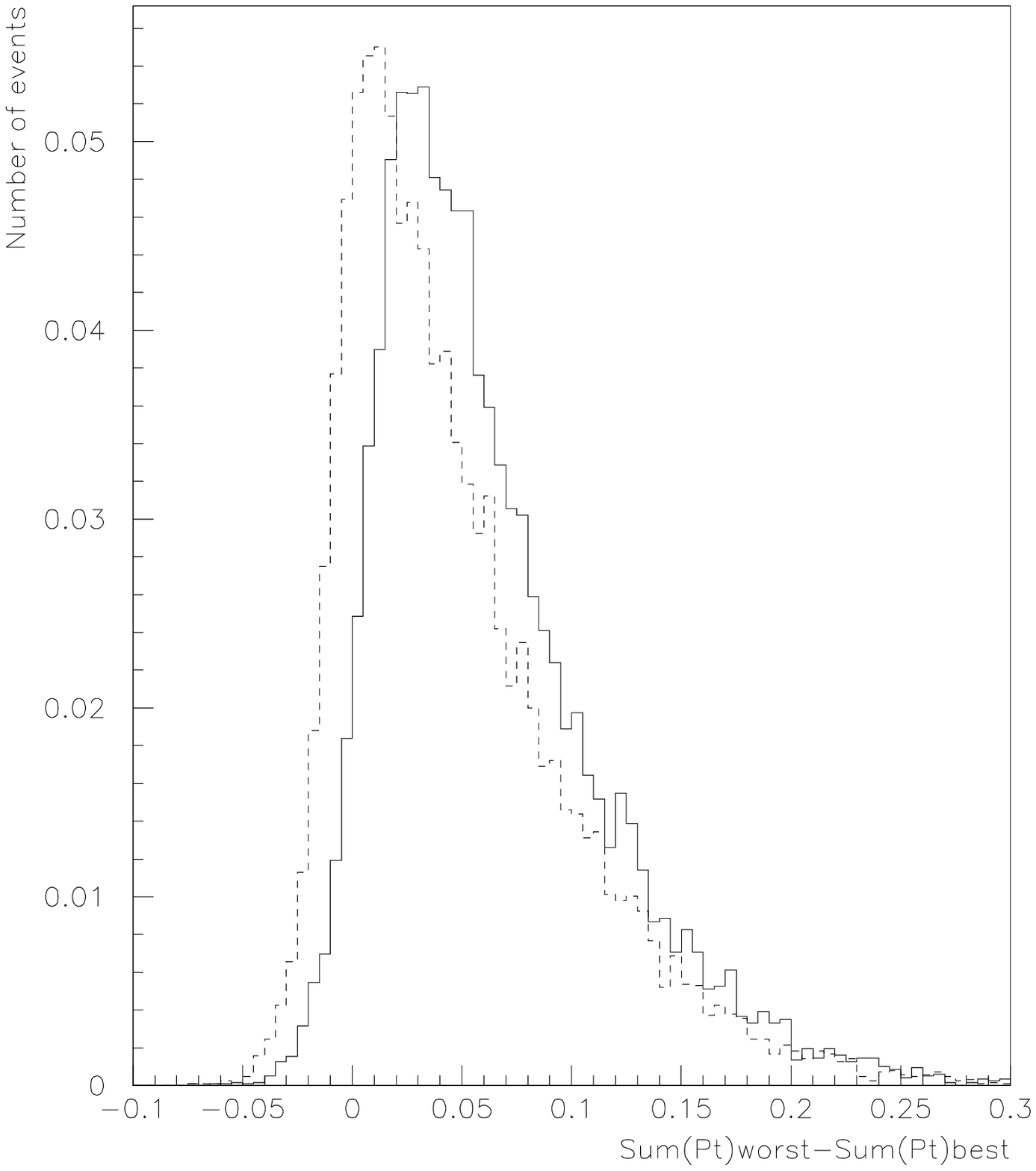, width=60mm, height=60mm}%
\hspace{10mm}
\begin{minipage}[b]{60mm}
\caption[]{Dashed lines are reconnected events according to model~2 ---
reconnection %
in origin after parton shower but before fragmentation. Full lines are ordinary
events and %
the distributions are:\\ %
\textbf{a}: \begin{math}(\sumpT)_{\mrm{best}}\end{math};\\ %
\textbf{b}: \begin{math}(\sumpT)_{\mrm{worst}}\end{math};\\ %
\textbf{c}: \begin{math}\Delta=(\sumpT)_{\mrm{worst}} - %
(\sumpT)_{\mrm{best}}\end{math}.}\label{fig:plot1}
\end{minipage}
\end{center}
\vspace{-5mm}\hspace{45mm}(c)
\end{figure}

Instead we want to construct a continuous observable whose distribution
can be studied. This could be done in the following way:
For every event, method~3 tells us which hypothesis
in figure~\ref{fig:4jet} is the best, second best, and the worst.
At he same time method~5 gives a value of $\sumpT$. Without reconnection
these methods should give the same result, and they often do, but if the strings
are allowed to reconnect the methods do not give the same result as often.
This is verified in table~\ref{table2}.
One possibility could then be to study the distribution of
$\sumpT$ for the best configuration according to method~3 above.
This curve should be shifted to larger $\sumpT$ if we reconnect the strings.
This has been done in figure~\ref{fig:plot1}(a) but the effects
are not very large. If, instead, we  plot $\sumpT$ for the distribution of the
worst
combination according to method~3 the curve is shifted in the other
direction (figure~\ref{fig:plot1}(b)).
This means that if we take the difference, $\Delta=(\sumpT)_{\mrm{worst}} -
(\sumpT)_{\mrm{best}}$,
the effect will be enhanced, as in figure~\ref{fig:plot1}. The terms `best' and
`worst'
are always referred to the prediction of method~3.

Some variants of this theme would be to study $\sum{\pT^2}$ or ${\sumpT/N}$ and
the
distribution of the corresponding differences. The first one exaggerates large
$\pT$ which is not really what we want. The other one normalizes the sum
so that the number of particles in the sum does not affect the result.
The result is about the same in any case so we stick with $\sumpT$ as our
variable.

\begin{figure}
\begin{center}
\mbox{\epsfig{file=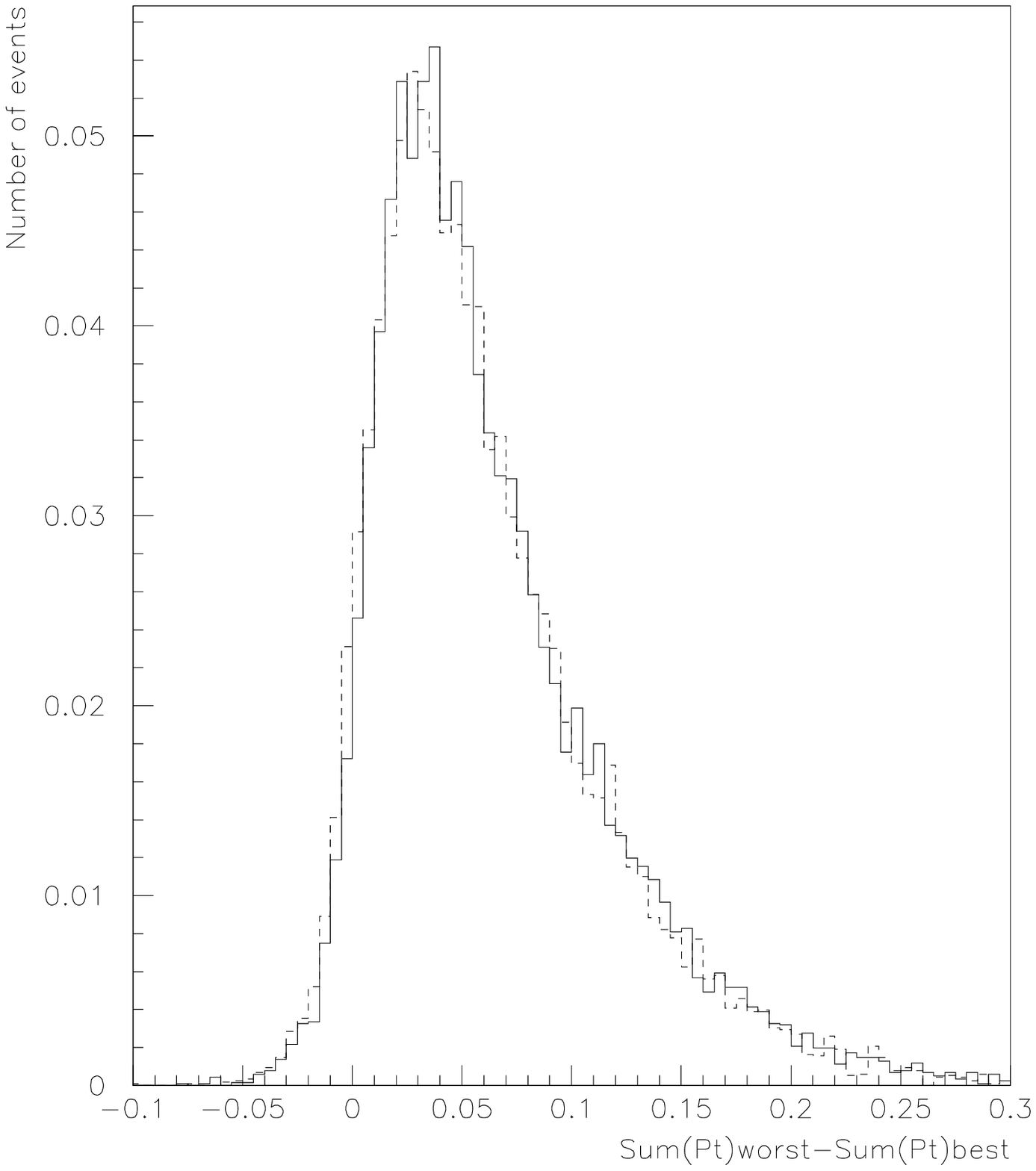, width=70mm, height=70mm}\hspace{10mm}
\epsfig{file=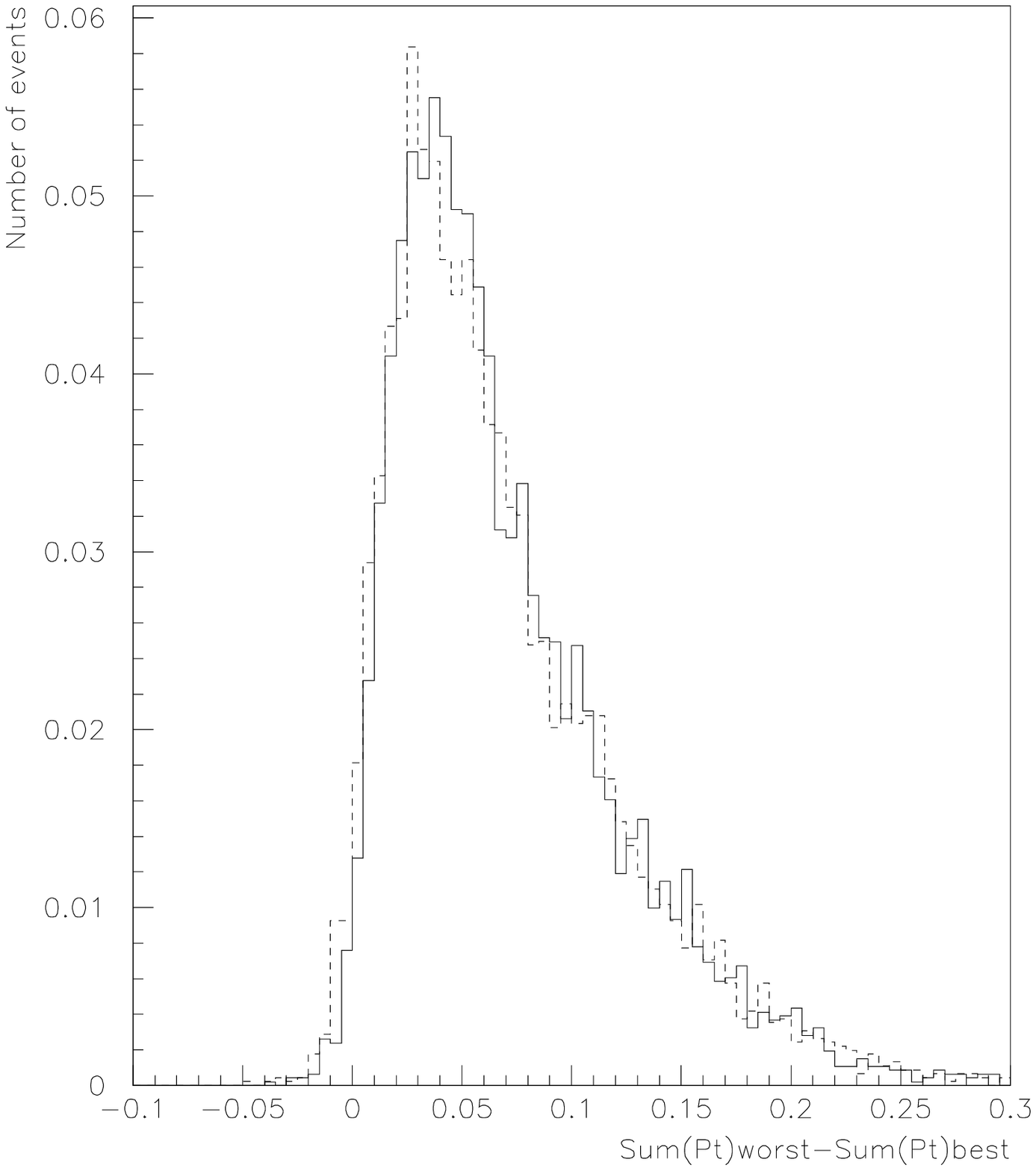, width=70mm, height=70mm}}
\mbox{(a)\hspace{70mm}(b)}
\end{center}
\caption[]{Distribution of $\Delta=(\sumpT)_{\mrm{worst}} - %
(\sumpT)_{\mrm{best}}$. %
Dashed lines are reconnected events according to the Type~I (a) and %
type~II (b) superconductor models. In the reconnection scenario not every %
event is reconnected, but only about 30\%.}\label{fig:plot2}
\end{figure}

Model~2 is neither very realistic nor appealing because it does not make any
assumptions about
the structure of the string. Model~3 and~4  predicts that the string behaves
like the vortex lines
of a Type I and type II superconductor respectively, and are two of the favorite
candidates in a description of the QCD vacuum~\cite{GPZ, SjoValery}.
Figure~\ref{fig:plot2} shows the same plot as figure~\ref{fig:plot1}(c) but here
the strings are reconnected according to models number~3 and~4 (type I and II
superconductor).
In these more realistic reconnection models not all events are reconnected. The
dashed `reconnected' events therefore contain mostly un reconnected events and
only about 30\% reconnected ones. This is a property inherent in the models,
because
events with and without reconnection does not look the same before the strings
are reconnected. In this case the effects
are clearly negligible, especially if we consider the statistics that is
available.
In the next section we will study some simpler toy models where we can study
where the effects diminish---effects that clearly exist at some level.

\subsection{Toy models}

\begin{figure}
\begin{center}
\mbox{\epsfig{file=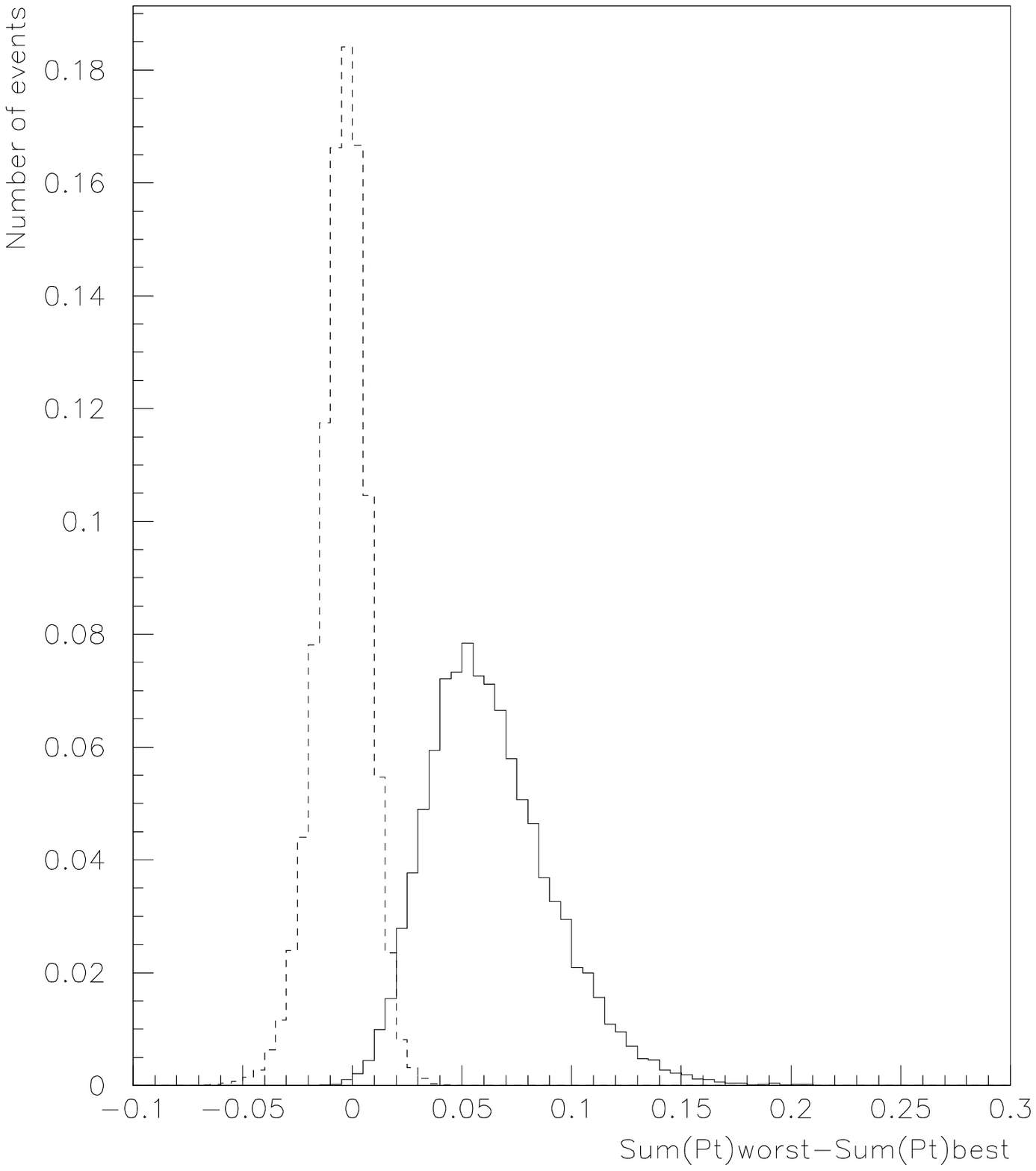, width=70mm, height=70mm}\hspace{10mm}
\epsfig{file=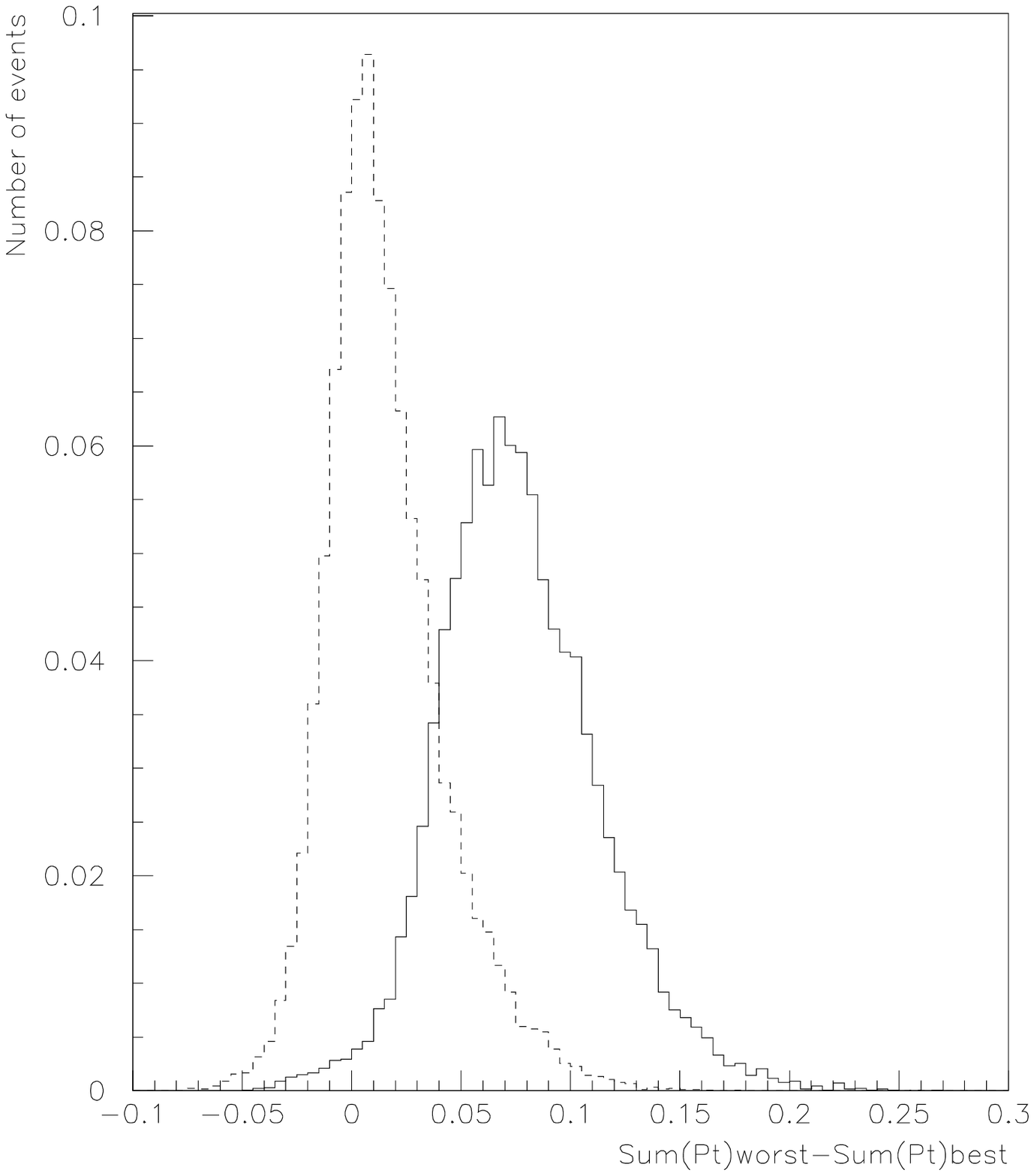, width=70mm, height=70mm}}
\mbox{(a)\hspace{70mm}(b)}
\caption[]{Ordinary and reconnected events according to toy models 1 and 2 (a %
and b respectively). %
Dashed lines are reconnected events. Fixed geometry.}\label{fig:toy1}
\end{center}
\end{figure}

\begin{figure}
\begin{center}
\mbox{\epsfig{file=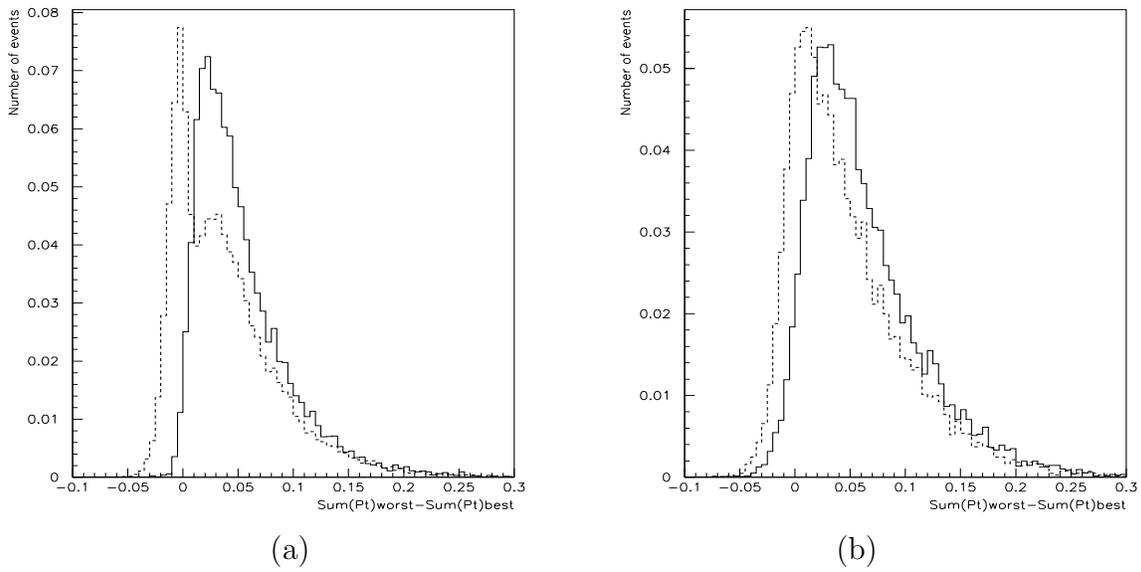, width=70mm, height=70mm}\hspace{10mm}
\epsfig{file=out4a.eps, width=70mm, height=70mm}}
\mbox{(a)\hspace{70mm}(b)}
\caption[]{Ordinary and reconnected events according to toy models 3 and 4 %
(a and b respectively). %
Dashed lines are reconnected events. %
No restrictions on the geometry except that the angle between any two jets is %
not too small.}\label{fig:toy2}
\end{center}
\end{figure}

Several toy models can be studied to see how the considered method
behaves under different simplified conditions:
\begin{Enumerate}
\item The most extreme example is a $\W^+\W^-$-pair produced at rest with a
center of mass energy of 160 GeV where the $\W$ particles decay to $\u\dbar$
and $\d\ubar$ quarks, and
the angle between the $\u$ and $\ubar$ quarks are $60^{\circ}$.
Parton showers are turned off and the strings are reconnected
at the origin before fragmentation.
\item Same as above but with parton showers. Reconnection after shower
but before fragmentation.
\item $\W^+\W^-$ at 170 GeV and free geometry. All quarks allowed but no
parton shower. Reconnection at origin before fragmentation.
\item Same as above but with showers. Reconnection after shower
but before fragmentation.
\end{Enumerate}
The distributions of $\Delta=(\sumpT)_{\mrm{worst}} - (\sumpT)_{\mrm{best}}$ in
these four cases are
depicted in figure~\ref{fig:toy1} and~\ref{fig:toy2}.

In these plots it is possible to study how the geometry and the parton showers
affect
the distribution of $\Delta$. It is clear that the geometry makes a big
difference --- 
compare plot (a) of figure~\ref{fig:toy1} and~\ref{fig:toy2} --- so if we
introduce even more restrictive cuts than those already presented some
improvements could
be made. The problem then is that we lose a lot of events and we will not be
able to
compare with experiment. The parton showers simply smears out the distribution
and generally makes it harder to see the differences.

The reconnected event sample in figure~\ref{fig:toy2}a shows an interesting
shape. It looks like
a combination of two distributions: one like \mbox{figure~\ref{fig:toy1}(a,
dashed)}
and one like \mbox{figure~\ref{fig:toy2}(a, full)}. Since the only difference
between
\ref{fig:toy1}a and \ref{fig:toy2}a is that in \ref{fig:toy2}a the geometry of
the
jets is not fixed, some other restrictions on the geometry than those already
imposed
could separate events where the difference between reconnected and ordinary
events
is large from those where it is small. In figure~\ref{fig:toy2}b the parton
shower
smears out the distributions
and the two bumps merge. If we could separate the two
distributions the difference should be noticeable even here. The problem is
to find a variable that distinguishes the different distributions.
\begin{figure}
\begin{center}
\begin{picture}(120,90)(0,0)
\LongArrow(60,60)(10,80)
\Text(0,90)[]{$\q_1$}
\LongArrow(60,60)(110,20)
\Text(120,10)[]{$\qbar_2$}
\LongArrow(60,60)(10,10)
\Text(0,0)[]{$\qbar_4$}
\LongArrow(60,60)(110,80)
\Text(120,90)[]{$\q_3$}
\DashLine(10,80)(110,20){4}
\DashLine(10,10)(110,80){4}
\Text(64,27)[]{$\theta$}
\Curve{(54,40.8)(64,36)(74,41.6)}
\end{picture}
\end{center}
\caption[]{Definition of the $\cost$ variable. %
The quarks come from the reactions \mbox{$\W^+\rightarrow\q_1\qbar_2$} %
and \mbox{$\W^-\rightarrow\q_3\qbar_4$}}\label{fig:anglevar}
\end{figure}

\begin{figure}
\begin{center}
\mbox{\epsfig{file=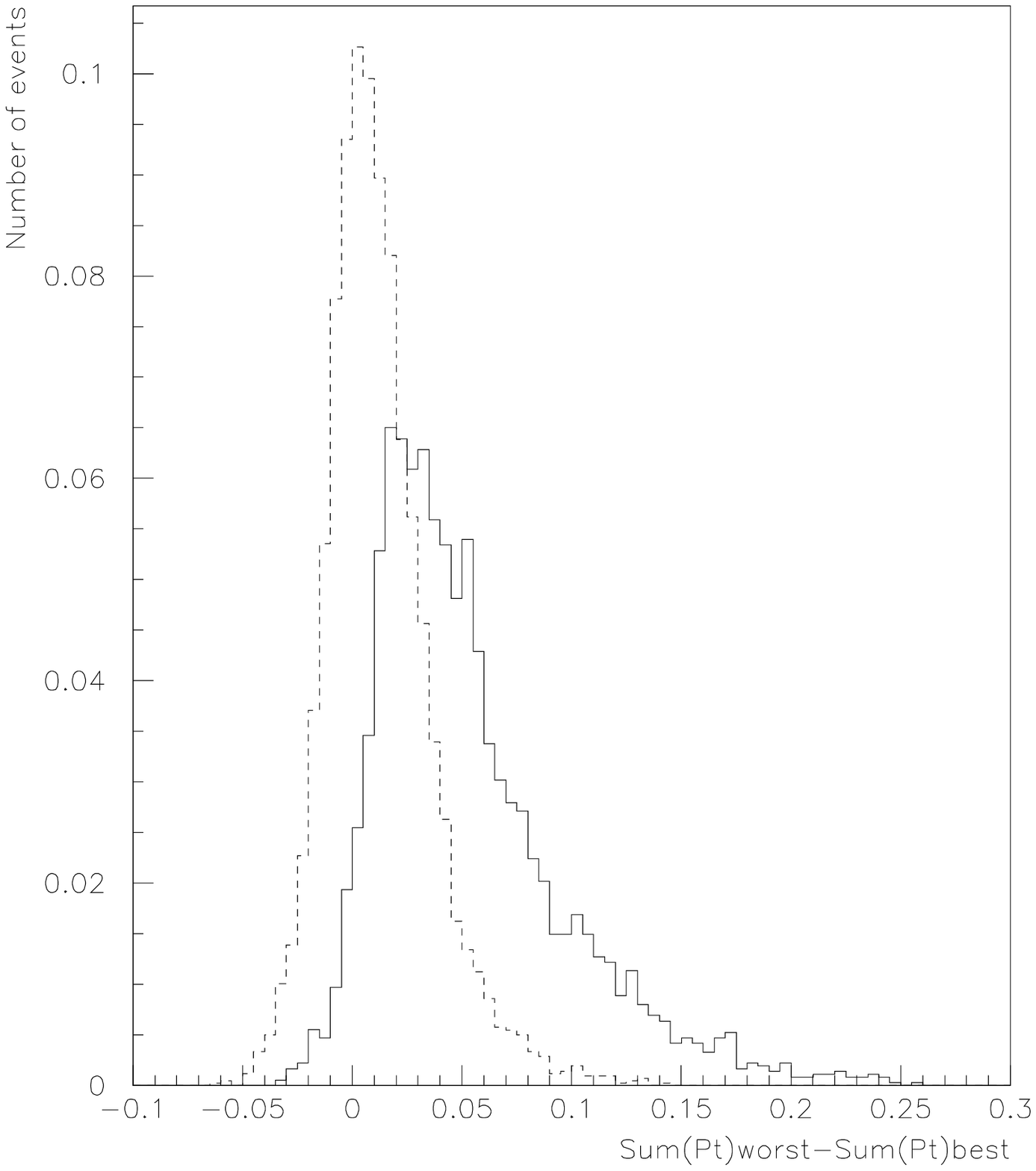, width=70mm, height=70mm}}\hspace{10mm}
\epsfig{file=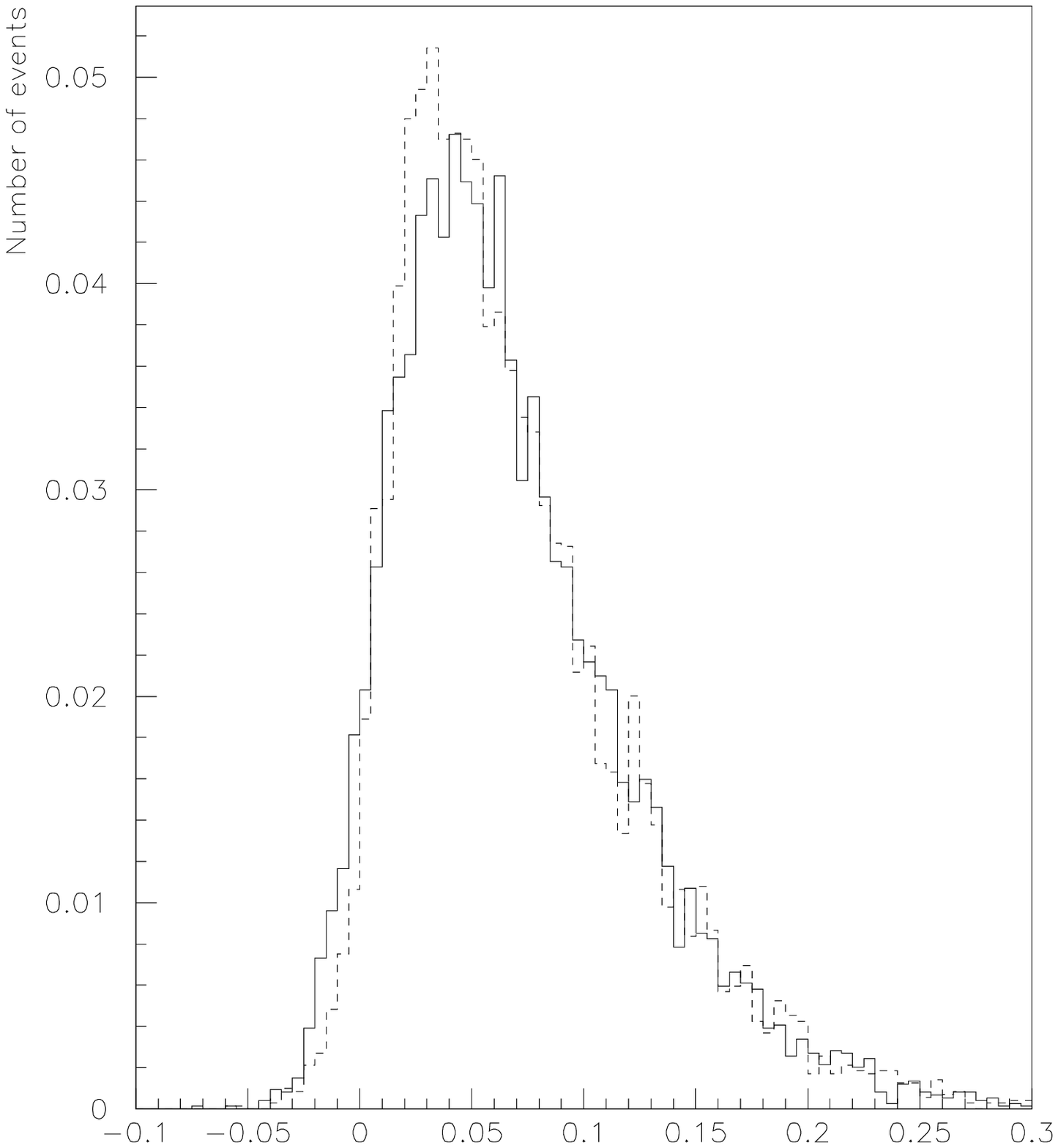, width=70mm, height=70mm}
\mbox{(a)\hspace{65mm}(b)}
\caption[]{Dashed lines are reconnected events according to toy model~4. %
Events with \textbf{a}: $\cost<0$ \textbf{b}: $\cost>0$. %
The difference between normal and reconnected events is much larger in %
a.}\label{fig:toy3}
\end{center}
\end{figure}

One possibility is the angle in figure~\ref{fig:anglevar}. This is
the angle between the direction vectors
$\nbar_1=\pbar_{\qbar_2}-\pbar_{\q_1}$ and
$\nbar_2=\pbar_{\qbar_4}-\pbar_{\q_3}$
and it has a value between zero and $\pi$.
Reconnected events have one string piece between $\q_1$ and $\qbar_4$
and one between $\qbar_2$ and $\q_3$. If the $\theta$ angle
is large the reconnected string pieces will be much smaller
than the original ones, which will make the two configurations more different.
If, on the other hand, the angle is small the string configuration
will be more similar to an event without reconnection.
In figure~\ref{fig:toy3} the distribution of figure~\ref{fig:toy2}b is separated
in
events with different signs of $\cost$. For $\cost>0$ ($\theta<\pi/2$, long
reconnected
strings) there is no noticeable difference
between reconnected and ordinary events. On the other hand, for events with
$\cost<0$
($\theta>\pi/2$, short reconnected strings), the
difference is clearly larger than in figure~\ref{fig:toy2}.

\begin{figure}
\begin{center}
\mbox{\epsfig{file=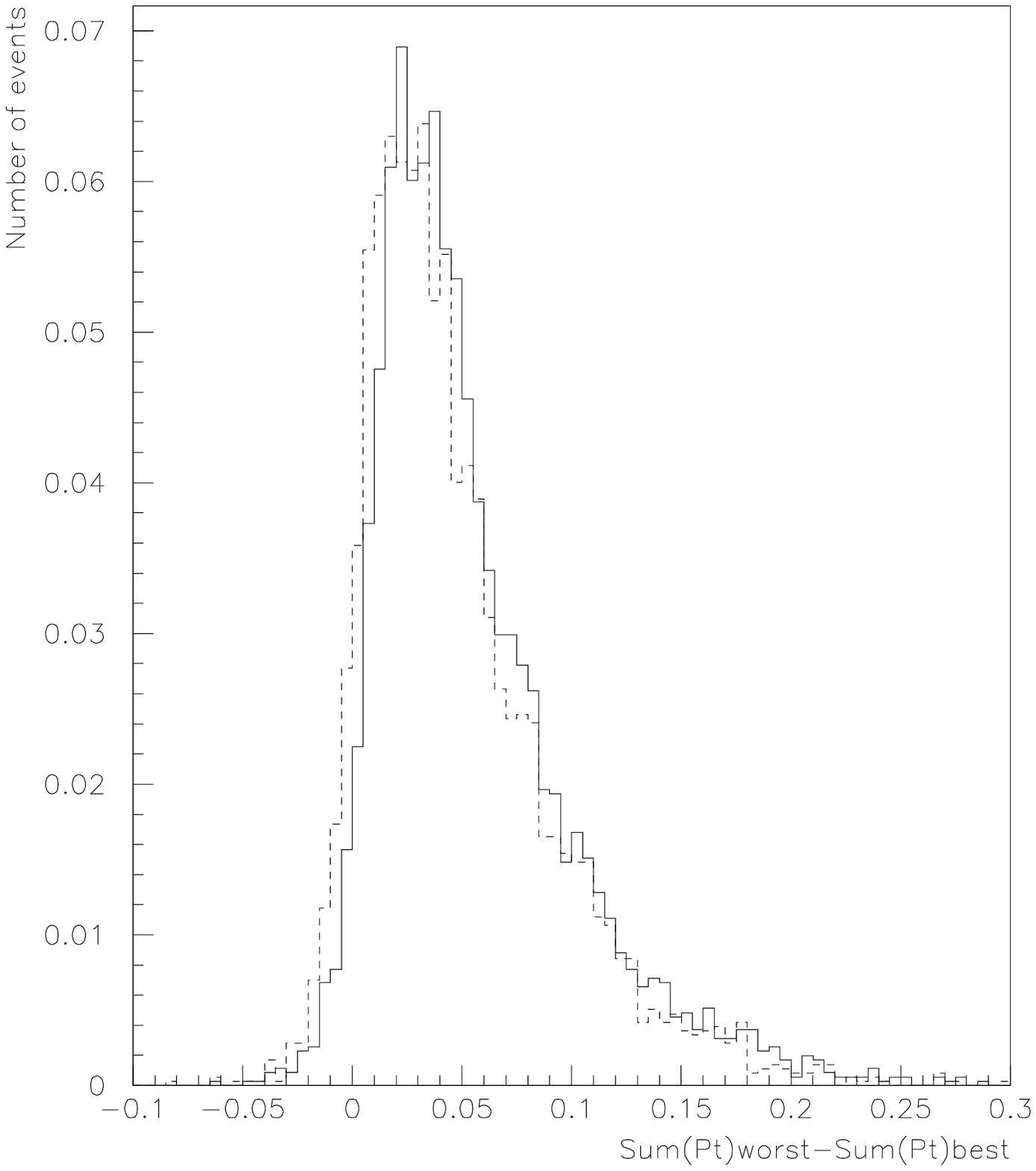, width=70mm, height=70mm}\hspace{10mm}
\epsfig{file=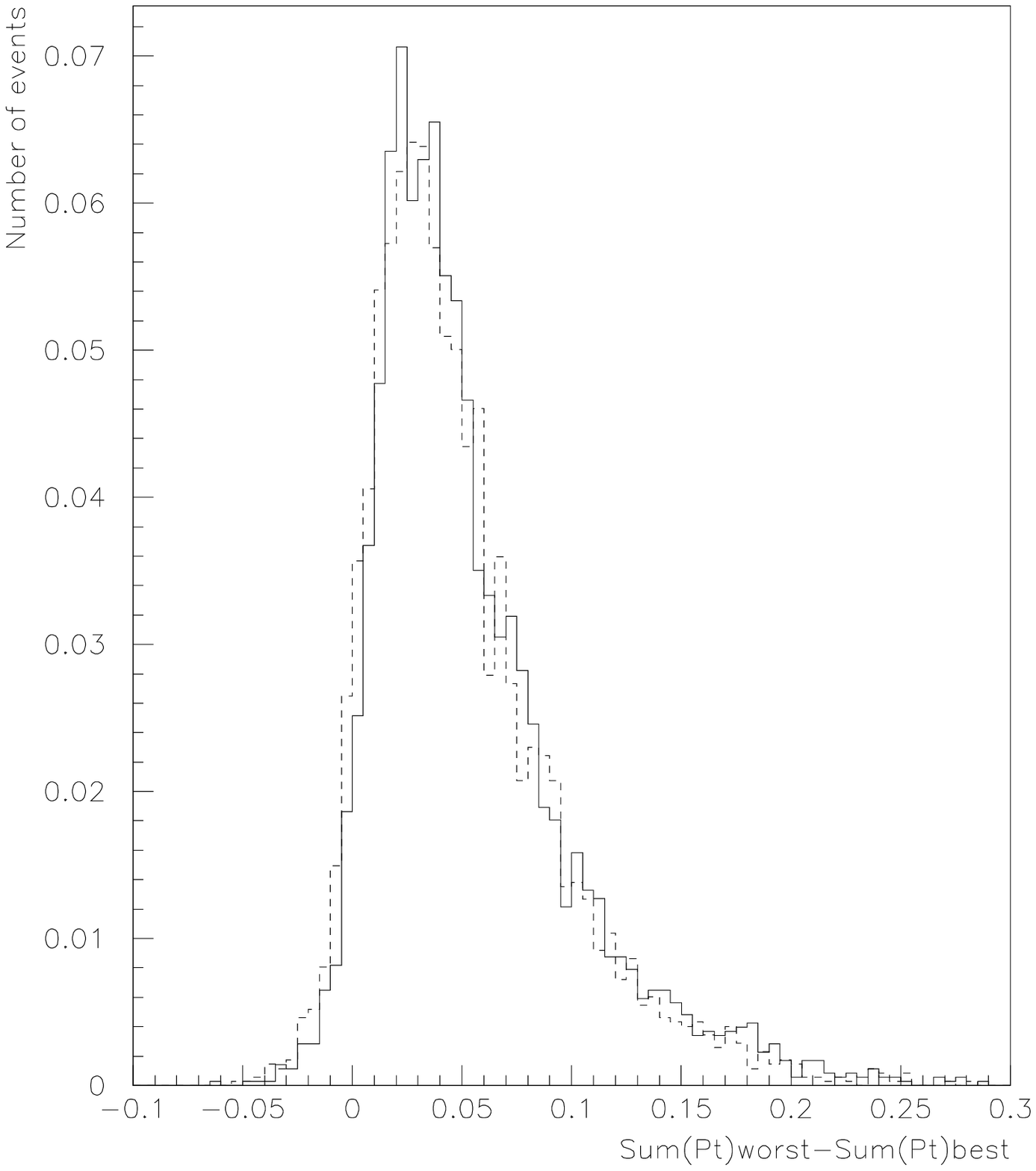, width=70mm, height=70mm}}
\mbox{(a)\hspace{65mm}(b)}
\caption[]{Dashed lines are reconnected events according to: \textbf{a} type~I
superconductor %
model \textbf{b} type~II superconductor model. The reconnection models contain %
about 30\% %
reconnected events. Only events with $\cost<0$ considered. %
20,000 events are generated and only about 3500 events survive all the %
cuts.}\label{fig:plot3}
\end{center}
\end{figure}

From the above we conclude that an additional cut could be to throw away all
events with $\cost>0$ for the more realistic models as well. This has been
done in figure~\ref{fig:plot3}. The signal is clearly larger than in
figure~\ref{fig:plot2} where there is hardly any difference between
reconnected and ordinary events. The problem here is that we have used
information
not available in an experimental situation, because it is not possible to say
directly which jet belongs to which quark. It is possible, however, to tag the
jets,
using the flavour and charge of the quarks (e.g. charm tagging), but
many events are lost in the process and we will be left with very poor
statistics indeed.
In the analysis made here 20,000 events producing two $\q\qbar$-pairs is
generated, out of these
10,500 survive the initial cuts and after the $\cost<0$ cut about 3,500 events
are left. This means
that about 20\% of the events are left for analysis. Scaling to LEP2 statistics
less than
one thousand events will be left and then the quark tagging remains to be done
as well.

\subsection{Other models}

\begin{figure}
\begin{center}
\mbox{\epsfig{file=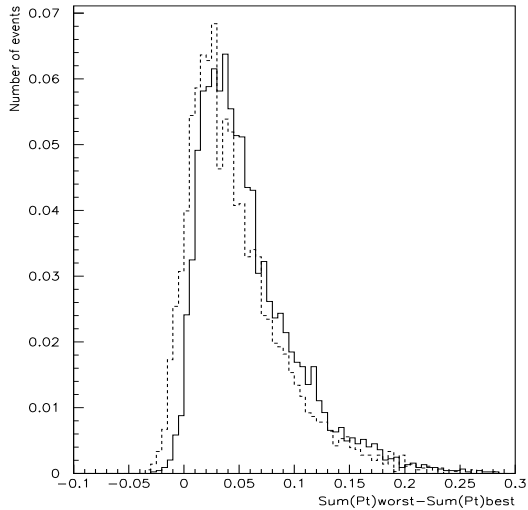, width=70mm, height=70mm}}
\caption[]{Reconnected events according model 5, %
the lambda measure minimization %
model. Dashed lines are reconnected events.}\label{fig:lambda1}
\end{center}
\end{figure}

A somewhat different model, introduced in~\cite{GJari} and briefly described in
section~\ref{sec:theory:reconnection}, has also been studied using the
$\sumpT$-method. Figure~\ref{fig:lambda1} shows the corresponding distribution
of $\Delta$ in this scenario. This model depends on the dynamics of the
strings and not on the string structure itself, consequently it does not address
the
issue of the structure of the QCD vacuum. Instead it is based on the principle
that the string length should be preferentially minimized by reconnection.
It is interesting to note that the effects here are quite large even without
the $\cost<0$ cut. It might be possible to compare this scenario with experiment
when the data is available. It should be noted, though, that all events
in the dashed plot is reconnected ones. In a real event the reconnection
probability could be much smaller.

\section{Summary and outlook}

It is reasonably clear by now that colour reconnection is an effect that exists
at some level but it is hard to say whether reconnection will be possible to
observe
at LEP2. Perturbatively the effect is negligible and in the fragmentation
region the phenomenon is not understood from first principles. Several models
have been studied, ranging from simple ones where strings are reconnected at
the origin to models where the string structure behaves like the
vortex lines of a superconductor.

In this work simple models show
large differences between ordinary and reconnected events,
whereas the more realistic ones do not. It has been shown that the method can,
in principle,
distinguish experimentally a sample with reconnected events from one with
ordinary
events. Because of this it will most probably be
possible to either reject or confirm the simpler models, for example the
instantaneous
and intermediate ones studied in~\cite{GPZ},
while the testability of the superconductor models is uncertain, due to the
small number
of experimental events that will be available for comparison. Even the model
in~\cite{GJari}, the lambda minimization reconnection scenario, is difficult
to test. If the experimental basis had
been larger, some cuts that we have proposed would have made it possible
to select interesting events where the geometrical difference between
ordinary and reconnected events are larger and hence will give a larger
difference
in the plots. Even with these cuts it has been shown that
the parton showers blurs out much of the effects.

The fact that the effects are small is not entirely a bad
thing since this means that there will be only small
systematic uncertainties in the measurement of the $\W$-mass. This uncertainty
has
already been approximated in~\cite{SjoValery} to be in the order of 40 MeV,
which
in itself is fairly small. On the other hand, the possibility to study
the QCD vacuum with this method seems to be small. In order to make any real
predictions the
differences have to be more pronounced, especially when we consider the
lack of experimental data. Therefore, as things stand today,
we have to wait for results from experiment to make further progress.

The $\pT$ method that has been introduced can be used to study string structure
in other cases than colour reconnection and two examples has been given:
three-jet
and four-jet events in $\Znoll$ decays where differences between independent
fragmentation and Lund string fragmentation can be studied. In three-jet events
the matter is pretty clear and the presence of the strings has been verified
in experiment. In four-jet events the situation is more complicated
because of the geometry and the many different possible ways to draw the
strings. A simple example has been given here and a possible method has been
described, in principle, but it needs to be more refined in order to be
realistic.\\[4mm]
{\bf acknowledgments}\\
I would like to thank my advisor for always having time for me, my roommates
for good company and the department for general comfort. Special thanks
goes to the graduate students who has always had time to help me with computer
related
problems. My girlfriend proofread the manuscript which improved the English a
lot. Still, any mistakes
left are my own.



\begin{thebibliography}{99}

\bibitem{Lundmod}
B. Andersson, G. Gustafson, G. Ingelman and T. Sj\"ostrand, \prep{97}{1983}{31}

\bibitem{Manual}
T. Sj\"ostrand, {\em Comput. Phys. Commun.} {\bf 82} (1994) 74

\bibitem{QCDEventGenerators}
I.G. Knowles {\em et al}. in `Physics at LEP2', eds. G. Altarelli, T.
Sj\"ostrand and F.~Zwirner,
CERN 96--01, Vol. 2, p.103

\bibitem{GPZ}
G. Gustafson, U. Pettersson and P. Zerwas, \pl{B209}{1988}{90}

\bibitem{SjoValery}
T. Sj\"ostrand and V.A. Khoze, \zp{C62}{1994}{281}

\bibitem{reconnectionsummary}
Z. Kunszt {\em et al.} in `Physics at LEP2', eds. G. Altarelli, T.~Sj\"ostrand
and F.~Zwirner,
CERN 96--01 vol. 1, p.141

\bibitem{GJari}
G. Gustafson and J. H\"akkinen, \zp{C64}{1994}{659}

\bibitem{crister}
C. Friberg, G. Gustafson and J. H\"akkinen, LU TP 96-10 (submitted to {\em
Nucl.~Phys.~B})

\end{thebibliography}
\end{document}